\documentclass[12pt]{article}
\usepackage{jheppub}
\usepackage[top=1in, bottom=1in, left=1in, right=1in]{geometry}
\usepackage[font=footnotesize]{caption}
\usepackage{array}
\usepackage{enumerate}
\usepackage{float}
\usepackage{subfig}
\usepackage{multicol}
\usepackage{multirow}
\usepackage{epsfig}
\usepackage{graphicx}
\usepackage{pict2e}
\usepackage{color}
\usepackage{amsmath,amsfonts,amssymb}
\usepackage{bbm}
\usepackage[utf8]{inputenc}
\usepackage[english]{babel}
\usepackage{xspace}

\newcommand{\be}{\begin{equation}}
\newcommand{\ee}{\end{equation}}
\newcommand{\bea}{\begin{eqnarray}}
\newcommand{\eea}{\end{eqnarray}}

\title{Generalized Clockwork Theory}

\begin{abstract}
{We generalize the clockwork theory in several directions. First, we consider beyond nearest neighbors interactions.  Considering such interactions keeps a larger subgroup of the original $U(1)^{N+1}$ unbroken and can allow for different symmetry breaking patterns.  We recover the original clockwork scenario in the presence of these additional interactions. In such case, the masses of the massive modes change, but a single massless mode remains intact. Such interactions are naturally interpreted as higher derivative terms from the point of view of extra dimensions. Second, we generalize the clockwork shift symmetry to non-abelian global groups. Third, trivial embedding of the clockwork scenario in supergravity, yields an AdS minimum as big as the clockwork interaction. Specifically, the clockwork is connected to the cosmological constant. We analyze the different ways in which a Minkowski supersymmetric minimum can be constructed, and demonstrate simple SUSY breaking mechanisms that preserve or break the clockwork symmetry. We show that the clockwork direction is actually a special SUSY breaking direction, that does not require the inclusion of additional fields or parameters.
 Fourth, we review the extra-dimensional origin of the mechanism and interpretation, in the case of conformal coupling to gravity. 
 }
\end{abstract}
\author{Ido Ben-Dayan}
\affiliation{Department of Physics,
Ben-Gurion University of the Negev, P.O. Box 653, Be'er-Sheva 8410500, Israel\\
Physics Department, Ariel University, Ariel 40700, Israel
}
\emailAdd{ido.bendayan@gmail.com}
\begin{document}

\maketitle

\section{Introduction}
Clockwork Theory (CW) has been proposed as a mechanism of generating light particles with suppressed interactions, while no small parameters exist in the UV theory, \cite{Kaplan:2015fuy,Choi:2015fiu,Giudice:2016yja}. 
An earlier incarnation is actually in the context of having a superplanckian axion decay constant in Natural Inflation models, in the case of many sites \cite{Choi:2014rja}, or simple two sites case, \cite{Kim:2004rp, Ben-Dayan:2014zsa, Tye:2014tja}. Given that a large portion of contemporary theoretical physics involves generation of small/large numbers from $O(1)$ numbers in a 'natural' way, several new applications of the idea have been suggested, like a clockwork WIMP \cite{Hambye:2016qkf}, clockwork Inflation \cite{Kehagias:2016kzt}, clockwork composite Higgs \cite{Ahmed:2016viu}, photophilic QCD axion \cite{Farina:2016tgd}, and a solution to the hierarchy problem, \cite{Giudice:2016yja}.

The basic framework considers $N+1$ complex scalar fields with global $U(1)^{N+1}$ symmetry. The symmetry is explicitly broken by 'nearest neighbors' interaction to a single $U(1)_{CW}$. Thus, we have a 'theory lattice', where each scalar is sitting on a different site and interacts only with its nearest neighbors. In such case, there is a single massless mode, where its overlap with the j-th site goes like $\sim q^{-j},\, q>1$. Hence, if we couple "matter fields" to the $N$-th site, the massless mode coupling to these matter fields is suppressed by $q^{-N}$.
The idea can be implemented for scalars, fermions, gauge fields and gravitons (at least at the linear level). In all cases a massless particle remains while the other $N$ have a rather dense mass spectrum, all around the fundamental scale of the theory. In \cite{Giudice:2016yja}, it was suggested that such a framework may come from deconstructing a $5D$ scalar, potentially coming from 'Little String Theory'. A recent analysis in \cite{Craig:2017cda} has shown that the CW phenomenon is purely abelian, and cannot come from purely geometric extra-dimensional effects. However, see the response \cite{Giudice:2017suc}. A rather interesting suggestion is discarding the use of elementary scalar fields and deriving a 'clockworked' behavior from a sequence of strongly coupled sectors \cite{Coy:2017yex}.

To avoid obscurity, our definition of the clockwork is the following: 
Considering a theory with $N+1$ fields, charged under a $\mathcal{G}^{N+1}$ symmetry group with charge $q$, a 'clockwork' term is turned on leaving a residual $\mathcal{G}$ symmetry and a massless mode, such that its overlap with the $N$ massive modes behaves as  $q^{-j}$ for the j-th field.

Beauty is in the eye of the beholder, and considering multiple fields to explain hierarchies is neither very elegant nor extremely new. The usefulness of CW lies in a natural generation of hierarchy, that goes as $\sim q^N$ rather than $\sim N$ in a theory whose fundamental parameters are of similar size. In this note, we offer several simple generalizations and observations that may be useful for Particle Physics phenomenology and Cosmology. Both areas riddled with hierarchy problems. 

The purpose of these generalizations is two-fold. First, understanding in what applications can we apply the CW mechanism in a useful way, that might be tested by observations.  Second, what is the UV theory that generates CW behavior.  

First, using effective field theory as a guiding principle, nothing forbids additional interactions of the CW mechanism beyond nearest neighbors interactions. We shall demonstrate that interactions of $k$ neighbors results in a residual symmetry group of $U(1)^k$. The low energy theory will have $k$ massless modes. These interactions will show up and modify the mass spectrum of the massive fields in the original CW scenario predicting a rather different phenomenology. For example, in the original CW scenario of complex scalar fields, the masses of the radial fields that get a vev $\sim f$ is parametrically larger than the massive axions $\sim \sqrt{\epsilon} f q$, where $\epsilon$ is a small coupling constant controlling the CW behavior and $q$ some small integer. By going beyond nearest neighbors, the axions now have a mass of $\sim \sqrt{\epsilon} f q^k$, where $k$ is the number of neighbors each site couples to. So in this setting, the axions can be more massive by a factor of $q^{k-1}$ compared to the simple CW model. However, for consistency, these masses are still lighter than the masses of the radial modes.  
Additionally, the SM and its extensions have multiple (sometimes anomalous) $U(1)$ global symmetries. By going beyond nearest neighbors we can accommodate such symmetries in a straight forward way.
  
  Considering the UV theory, we show that these beyond nearest neighbors interactions correspond to higher order derivatives in a 5D picture in section $3$. In section $4$, we generalize the CW mechanism to non-abelian global groups, and specifically to the $O(N)$ vector model.

Second, the CW has been realized in the supersymmetric context in \cite{Kaplan:2015fuy}. The idea is to get CW with SUSY vacuum, from which one can start model building by breaking SUSY. We discuss the supersymmetric CW in section $5$,  and try to embed it in supergravity (SUGRA). The SUGRA behavior turns out to be very different than the global SUSY case. An immediate generalization to supergravity (SUGRA) with canonical K\"ahler potential, preserves the $U(1)_{CW}$, but gives vevs to additional fields, yielding an AdS supesymmetric vacuum. The AdS minimum is controlled by the CW coupling. The vacuum is lifted to a Minkowski SUSY vacuum by adding a constant term to the superpotential.  

 Alternatively, using a stabilizer field leads to $N$ flat directions or CW that its energy scale is parametrically the cosmological constant. Instead, we offer a shift symmetric CW superpotential such that the resulting F-term scalar potential has the CW form with SUSY Minkowski minimum.
We then demonstrate how to break SUSY to a dS without breaking $U(1)_{CW}$ by introducing a spurion superfield. Furthermore, we show that breaking SUSY within the CW sector forces the massless CW mode to be the SUSY breaking direction. Then, no residual $U(1)_{CW}$ is left. Avoiding this conclusion requires the inclusion of additional parameters and is transparent once we move to the mass basis. 


Third,
to diminish the arbitrariness in invoking $N+1$ scalar fields, one can view the CW as discretizing an extra dimension \cite{Giudice:2016yja}.
In the continuum limit, the theory behaves as a linear dilaton coming from Little String Theory, \cite{Antoniadis:2011qw,Baryakhtar:2012wj,Cox:2012ee}. 
This allows the extra dimension to be warped. However, contrary to the Randall-Sundrum case where the warping is exponential, here the warping is polynomial. Thus, for the correct Planck mass, and a new physics scale at $10\, TeV$ the proper size of the extra dimension is considerably larger, at the nanometer level \cite{Cox:2012ee}.
Up to now, the residual CW $U(1)$ was explicitly broken by some operator, usually coupling the $N$th field to some different sector of the theory, for example the SM making the framework "technically natural". It is interesting to consider whether the explicit breaking of the CW symmetry can arise naturally, rather than adding it by hand according to the problem one wishes to solve. In such case the theory will not be "technically natural", but simply "natural" since all the couplings and scales will be determined by the UV theory. 
Therefore, in section $6$, after reviewing the 5D picture we conformally couple the 5D free scalar field to the Ricci scalar and/or add a mass term, that in 4D will provide an explicit breaking term to the CW symmetry. Upon discretization the coupling to the Ricci scalar makes the massless mode tachyonic, while the 5D mass term obviously gives a positive mass to the massless mode. We suggest a conformal coupling of the clockwork to a positively curved 5D manifold. This time, the conformal coupling to gravity gives a positive mass term to the massless CW mode. We find similar mass scales, but now the CW charges $q$ become site-dependent, $q_j$. We analyze the case of site dependent charges in the Appendix. In both cases above, we assumed some other dynamics or energy sources, fixing the 5D metric. 

The outcome of these simple generalizations is a mixed blessing. On the one hand, from a lattice point of view the CW idea can be generalized to any number of neighbors and to global non-abelian symmetry groups. On the other hand, our investigation shows that CW is a rather delicate phenomena, and its embedding in SUGRA or beyond a free scalar field in $5D$ is problematic. The UV origin of CW theory is therefore obscure.
\section{Generalization of the Clockwork Mechanism}
\subsection{Review of the Clockwork Mechanism}
The original clockwork considers $N+1$ complex scalars, $\phi_j$, where $j=0,1,\cdots N$ with canonical kinetic terms, and a potential:
\be
\label{eq:original}
V(\phi_j)=\sum_{j=0}^N\left(-\tilde m^2\phi_j^{\dagger}\phi_j+\frac{\lambda}{4}|\phi_j^{\dagger}\phi_j|^2\right)+\sum_{j=0}^{N-1}\left(\epsilon \phi_j^{\dagger}\phi_{j+1}^3+h.c\right)
\ee
When $\epsilon \rightarrow0$ we have a global $U(1)^{N+1}$ symmetry. Turning on $\epsilon\ll \lambda<1$ breaks this symmetry to a $U(1)$.
Under the remaining $U(1)$, the fields have charges $3^{-j}$.
We expand around the vacuum of
the spontaneously broken theory $\langle |\phi_j|^2\rangle= f^2\equiv 2\tilde m^2/\lambda \, ,\forall j$. Below the breaking scale $\sqrt{\lambda}f$, we have a theory of $N+1$ goldstone bosons with $U_j=e^{i \pi_j(x)/f}$ and $j=0,\cdots N$: 
\be
\label{eq:lag1}
\mathcal{L}=-\sum_{j=0}^{N}f^2\partial U_j^{\dagger}\partial U_j+\frac{m^2 f^2}{2}\sum_{j=0}^{N-1}\left( U^{\dagger}_jU^q_{j+1}+h.c\right)
\ee
\eqref{eq:original} corresponds to $q=3$, but the $U$s are dimensionless, so we can consider any $q$, as is done in \cite{Giudice:2016yja}, and $m^2=2\epsilon f^2$. In general, one can assign different masses $m_j$ and different charges $q_j$ to each site. In terms of the low energy effective theory of the pions we have canonical kinetic terms and the following potential:
\be
V(\pi_j)=\frac{m^2}{2}\sum_{j=0}^{N-1}(\pi_j-q\pi_{j+1})^2+\mathcal{O}(\pi^4)=\frac{1}{2}\sum_{i,j=0}^{N}\pi_i M^2_{ij}\pi_j+\mathcal{O}(\pi^4)
\ee
The theory is invariant under the shift symmetry $\pi_j\rightarrow\pi_j+r/q^j$, where $r\in \mathbb{R}$. 


The mass matrix $M_\pi^2$ is given by
\be
M^2_\pi = m^2
\begin{pmatrix}
1 & -q & 0 & \cdots &  & 0 \cr
-q & 1+q^2 & -q & \cdots &  & 0 \cr
0 & -q & 1+q^2 & \cdots & & 0 \cr
\vdots & \vdots & \vdots & \ddots & &\vdots \cr
 & & & & 1+q^2 & -q \cr
 0 & 0 & 0 &\cdots & -q & q^2
\end{pmatrix} \, .
\label{pimass}
 \ee
Diagonalizing the mass matrix gives one massless mode and $N$ massive modes with successive mass splittings:
\be
\label{massa}
m^2_{a_0}=0, \quad m^2_{a_k}=m^2\lambda_k, \quad \lambda_k=\left(1+q^2-2 q \cos \frac{k \pi}{N+1}\right), \quad k=1,\cdots N
\ee
where the interaction basis $\pi_j$ and mass basis $a_j$ are related by:
\be
\pi=\mathcal{O}a,\quad \mathcal{O}^TM_{\pi}^2\mathcal{O}=diag(m^2_{a_0},\cdots m^2_{a_N})
\ee
 The rotation matrix and normalization are given by:
\be
\mathcal{O}_{j 0} = \frac{{\cal N}_0}{q^j} \, , ~~~ 
\mathcal{O}_{j k} = {\cal N}_k \left[ q \sin \frac{j k\pi}{N\! +\! 1}-  \sin \frac{(j +1) k\pi}{N\! +\! 1} \right] \, ,~~~ j =0, .. , N;~~k =1, .. , N
\label{rotation}
\ee
\be
{\cal N}_0 \equiv \sqrt{\frac{q^2-1}{q^2-q^{-2N}}} \, , ~~~~ {\cal N}_k \equiv \sqrt{\frac{2}{(N\! +\! 1)\lambda_k}} \, .
\ee

$\mathcal{O}_{j0}$ is the amount the of the massless mode $a_0$ contained in each pion $\pi_j$. Because $\mathcal{O}_{j0}\sim q^{-j}$ it becomes smaller by a factor of $q$ as we move away in $j$. Thus, the overlap with the last site, the Nth one, is exponentially suppressed. 
By coupling a theory like the Standard Model to the Nth site, we get an exponentially enhanced decay constant for the Goldstone interactions (with $a_0$), i.e. a scale exponentially larger than the scale of spontaneous symmetry breaking $f$:

\be
\mathcal{L}= \frac{\pi_N}{16\pi^2 f}F_{\mu \nu}\tilde F^{\mu \nu}
\Rightarrow\frac{1}{16\pi^2}F_{\mu \nu}\tilde F^{\mu \nu}\left(\frac{a_0}{f_0}-\sum_{k=1}^N\, (-1)^k\frac{a_k}{f_k}\right),\quad
 f_0=\frac{q^Nf}{\mathcal{N}_0}, \quad f_k=\frac{f}{\mathcal{N}_k q \sin \frac{k \pi}{N+1}}.
\ee

To summarize, in the original CW model, before coupling to the sector to the SM, one has a residual $U(1)$, a massless goldstone mode with 
eigenvector $\mathcal{O}_{j0}\sim q^{-j}$, massive axions with masses $m^2_{a_k}\sim \epsilon f^2 q^2$ and massive radial modes with masses $m_{r_k}^2\sim f^2$, (or $m_{r_k}^2\sim f^2(1+\epsilon)$ to be precise), parametrically heavier than the axions. Both the masses of the radial modes and axions are densely spaced. The above analysis requires the hierarchy of couplings, $\epsilon \ll \lambda \ll 1$. We shall see that this hierarchy is relaxed once we generalize to beyond nearest neighbors. 
\subsection{Generalization of the Clockwork Mechanism Beyond Nearest Neighbors}
  As an effective field theory, there is no reason to limit ourselves to nearest neighbors interactions as in \eqref{eq:lag1}, since many additional interactions still respect the $U(1)$ symmetry. One must take these interactions into account in a consistent way. \footnote{If we wish to follow the original discussion of a renormalizable theory in \eqref{eq:original}, we can still consider tree level interactions of the sort $\phi_j^{\dagger}\phi_{j+1}\phi_{j+2}^2+h.c$. Upon diagonalization, we will still have two massless modes with the overlap of the last site behaving as $2^{-(N-1)}$ and $2^{-(N-2)}$.}
 We expect these additional interactions to modify the resulting phenomenology. Indeed, the coupling to beyond nearest neighbors changes the symmetry structure of the theory and its spectrum in an interesting manner.
 
 Let us start by considering interactions between each site to the next to nearest neighbors. The potential with such couplings will look like:
 \be
 \label{eq:qp}
\frac{m^2 f^2}{2}\sum_{j=0}^{N-2}\left( U^{\dagger}_jU^q_{j+1}U^p_{j+2}+h.c\right)
 \ee
 Notice that since we couple each site to the two consecutive sites, we have to truncate the sum at $N-2$ instead of $N-1$.
Considering again the pions gives:
\be
V(\pi_j)=\frac{m^2}{2}\sum_{j=0}^{N-2}(\pi_j-q\pi_{j+1}-p \pi_{j+2})^2+\mathcal{O}(\pi^4)
\ee
Each term in the potential, is invariant under the transformation $\pi_j\rightarrow \pi_j+\alpha_j$ if:
\be
\alpha_j=2\pi \ell+q\alpha_{j+1}+p\alpha_{j+2}
\ee
 for integer $j$, and integer $\ell$. The original clockwork idea is the particular case of $p=0$. Obviously, we have here a two dimensional space, spanned by $\alpha_{j+1},\alpha_{j+2}$ and as expected, we have $U(1)^2$ residual symmetry, instead of $U(1)$. We also have two massless modes.
To recover the clockwork behavior we choose $q=\tilde q/2$ and $p=\tilde q^2/2$, then the potential will be made of terms $V\supset\left(\pi_j-\frac{\tilde q}{2}\pi_{j+1}-\frac{\tilde q^2}{2} \pi_{j+2}\right)^2$. Let us drop the tildes. In such case the original clockwork symmetry is conserved:
$\pi_j\rightarrow \pi_j+1/q^j$, and there is an additional $U(1)$ symmetry. 
\be
V(\pi_j)=\frac{m^2}{2}\sum_{j=0}^{N-2}\left(\pi_j-\frac{q}{2}\pi_{j+1}-\frac{q^2}{2} \pi_{j+2}\right)^2+\mathcal{O}(\pi^4)
\ee
Diagonalizing the mass matrix will give two massless states, one of which is the clockwork with $\mathcal{O}_{j0}=\mathcal{N}_0/q^j$. 
A straightforward check shows that a second massless eigenstate is given by an alternating vector:  $\mathcal{O}_{j1}\sim 2^j/(-q)^j$. However, such a vector is not orthogonal to the original clockwork mode. We orthogonalize the system using the Gram-Schmidt procedure that gives (no summation):
\be
\label{oj1ortho}
\mathcal{O}_{j1}=\frac{\mathcal{N}_1}{q^j}\left[(-2)^j-\frac{\sum_{i=0}^N\left(-2q^{-2}\right)^i}{\sum_{i=0}^Nq^{-2i}}\right]
\ee

 There are $N-1$ massive states that can also be brought to the desired orthonormal form by the Gram-Schmidt procedure. Their mass will be dominated by $q^4$ terms rather than $q^2$ in the original clockwork. The exact expression for the masses and eigenvectors can be obtained by recursion relations. However, it requires the analytical solution of a fourth order polynomial. While such a solution exists, it is not illuminating to write it down.
Coupling more and more neighbors will generate a polynomial of degree larger than four, that does not have an analytical general solution.  
Even without an analytical expression for the massive modes, the massless mode and the essence of the CW mechanism exists with the $q^{-j}$ overlap.

The generalization to any number of neighbors interactions is straightforward. Considering $n$ nearest neighbors interactions, the lagrangian will look like:
\be
\label{eq:anynumberfraction}
\mathcal{L}=-f^2\sum_{j=0}^{N}\partial U_j^{\dagger}\partial U_j+\frac{m^2 f^2}{2}\sum_{j=0}^{N-n}\left(U^{\dagger}_jU_{j+1}^{q/n}\cdots U_{j+n}^{q^n/n}+h.c\right)
\ee
For $n$ nearest neighbors interactions we will preserve $n$ symmetries and the conserved symmetry group will be some $U(1)^n$. Such a generalization allows for various breaking patterns, not necessarily reaching the $U(1)$ of the clockwork type. If we wish to maintain the same clockwork behavior of $\alpha_{j+1}/\alpha_j\sim q^{-1}$, then coupling to further neighbors makes the other massless eigenstates expression cumbersome.
It requires the simultaneous solution of:
\be
\alpha_j=\sum_{k=1}^n\alpha_{j+k}\frac{q^{j+k}}{n}, \quad
\frac{\alpha_{j+1}}{\alpha_j}=const.
\ee
where $n$ is the number of neighbors that are coupled.
Such recursive equations generate higher and higher polynomial equation for $\alpha_0,\alpha_1$, that for $n\geq6$ do not have a general analytic solution.

Analyzing the spectrum of the theory, we see a qualitatively different behavior. For $n$ nearest neighbors interactions, we have $U(1)^n$ symmetry, $n$ massless modes, and a different tower of massive modes. While the radial modes maintain their mass spectrum of $m_{r_k}^2\sim f^2$, the axions' mass spectrum behaves as $m^2_{a_k}\sim \epsilon f^2 q^{2n}/n^2$. Thus, depending on $q,n$, the axions can be parametrically heavier than the simplest CW model. 

Alternatively, if we wish to write the most general lagrangian that preserves only the original clockwork symmetry, we can add all possible neighbors interactions of this type. Thus, the full clockwork lagrangian is actually:
\be
\mathcal{L}=-f^2\sum_{j=0}^{N}\partial U_j^{\dagger}\partial U_j+\frac{m^2 f^2}{2}\sum_{k=1}^N\sum_{j=0}^{N-k}\left(U^{\dagger}_jU_{j+1}^{q/k}\cdots U_{j+k}^{q^k/k}+h.c\right)
\ee
In this lagrangian all the $U(1)$ symmetries except the original clockwork are broken, and a massless mode still remains and the component at each successive site remains $\mathcal{O}_{j0}=\mathcal{N}_0/q^j$. The masses of the radial modes remain $m_{r_k}^2\sim f^2$. The axions will be heavier, the dominant mass contribution $m^2_{a_k}\sim \epsilon f^2 q^{2N}/N^2$, pushing their masses towards the radial modes. 
Such different mass splitting could have interesting phenomenological consequences, since it changes model building scenarios such as the photophilic QCD axion of \cite{Farina:2016tgd}. 

The low energy lagrangian \eqref{eq:anynumberfraction} might seem questionable from a UV point of view, as it involves fractional charges. However, such form is quite abundant in type IIB string theory constructions \cite{Kachru:2003aw,BlancoPillado:2004ns,Grimm:2007hs,Blumenhagen:2006ci,Ben-Dayan:2013fva,Ben-Dayan:2014lca}.  For example, such potential
arises from D1 or D3 instantons wrapped around cycles in the Calabi-Yau manifold. A superpotential of the form
\be
W_{D3}=\sum_{\alpha} A_{\alpha}e^{-iJ^{\beta}_{\alpha}T_{\beta}}
\ee
where $T_{\beta}$ are moduli fields, $A_{\alpha}$ are coefficients that generically depend (weakly) on other fields in the spectrum, so are approximately constant, and $J_{\alpha}^{\beta}$ is a constant matrix parameterizing the warping numbers of the D3 instantons. In many known scenarios \cite{Kachru:2003aw,BlancoPillado:2004ns,Grimm:2007hs,Blumenhagen:2006ci,Ben-Dayan:2013fva,Ben-Dayan:2014lca}, the entries of $J_{\alpha}^{\beta}$ are of the form $\frac{2\pi}{N}$ where $N$ is a rank of a gauge group. Since $\frac{2\pi}{N}$ is an irrational number, it will give fractional charges. After stabilizing the imaginary part of $T_{\beta}$ the outcome are CW terms $W\sim e^{-i J^{\beta}_{\alpha}\Re{T_{\beta}}}$, such that $J^{\beta}_{\alpha}$ produces the CW form of \eqref{eq:anynumber}.
 Since $|W|^2\sim \cos (J^{\beta}_{\alpha}\Re{T_{\beta}})$, so $J^{\beta}_{\alpha}$ can be fractional and give fractional charges, and in particular $q^n/n$.
 
Another way to get fractional charges is by considering the 5D picture, for example in the case of a positively curved manifold in section $6$. The expression for the charge there is fractional and site dependent with
$q_j=\frac{\cos \left(\sqrt{3}/2 k j a\right)}{ \cos\left( \sqrt{3}/2 k (j+1) a\right)}$, where $a$ is the lattice spacing and $k$ is the CW spring constant.

\subsection{Clockwork Theory with Integer Charges}
Let us discuss a different possibility of getting a CW behavior beyond nearest neighbors, but with integer charges.
Generically, coupling beyond nearest neighbors interaction implies a nonrenormalizeable theory. Nonrenormalizeability also occurs in the original CW for $q>3$. Let us start by analyzing this case: 
Below the cutoff scale $\Lambda$,  the general potential looks like:
 \be
\label{eq:nonren}
V(\phi_j)=\sum_{j=0}^N\left(-\tilde m^2\phi_j^{\dagger}\phi_j+\frac{\lambda}{4}|\phi_j^{\dagger}\phi_j|^2\right)+\sum_{j=0}^{N-1}\left(\epsilon \frac{\phi_j^{\dagger}\phi_{j+1}^{q}}{\Lambda^{q-3}}+h.c\right)
\ee
Let us check whether previous analysis with the separation between the radial and axial modes still holds.
We would like to give each radial field an approximate vev as before of $\langle |\phi_j|^2\rangle= f^2\equiv 2\tilde m^2/\lambda \, ,\forall j$, as well as $f<\Lambda$. For this to happen, we need:
\be
\epsilon\frac{f^{1+q}}{\Lambda^{q-3}} q^2 \ll \lambda f^4 \Rightarrow \epsilon q^2 \ll \lambda \left(\frac{\Lambda}{f}\right)^{q-3}
\ee
Thus, for $q>3$, the desired hierarchy is easier to fulfill than the original CW. This will also be true when we couple beyond nearest neighbors. The masses of the radial modes are negligibly shifted to $m_{r_k}^2\sim f^2(1+\epsilon q^2(f/\Lambda)^{q-3})\sim f^2$.

Below the breaking scale $\sqrt{\lambda}f$, we have a theory of $N+1$ goldstone bosons with $U_j=e^{i \pi_j(x)/f}$ and $j=0,\cdots N$: 
\be
\mathcal{L}=-\sum_{j=0}^{N}f^2\partial U_j^{\dagger}\partial U_j+\frac{m^2}{2}\frac{f^{q-1}}{\Lambda^{q-3}}\sum_{j=0}^{N-1}\left( U^{\dagger}_jU^{q}_{j+1}+h.c\right)
\ee
with $m^2=2\epsilon f^2$.
The CW massless mode will now have the following eigenvector:
\be
\mathcal{O}_{j0}=\mathcal{N}_0\left\{1,\frac{1}{q},\frac{1}{q^2},\cdots,\frac{1}{q^N}\right\}
\ee
The masses of the axial modes behave as $m_{a_k}^2\sim \epsilon f^2(f/\Lambda)^{q-3}q^2$. The exact diagonalization is straightforward.
Apart from the massless CW mode, what kind of mass hierarchy can we get between the radial and axial modes?
The radial modes have masses of $m^2_{r_k}\sim f^2$ as long as $\epsilon q^2 \ll \lambda \left(\frac{\Lambda}{f}\right)^{q-3}$. The mass of the axions is $m^2_{a_k}\sim \epsilon f^2(f/\Lambda)^{q-3}q^2$. The 'standard hierarchy' is then $m_{r_k}^2>m_{a_k}^2$, which corresponds to $1\gg  \epsilon (f/\Lambda)^{q-3}q^2$ that is trivially satisfied since $1>\lambda \gg \epsilon q^2(f/\Lambda)^{q-3}$.

Let us generalize the CW nonrenormalizable model by considering interactions between each site to the next to nearest neighbors. The potential with such couplings will look like:
\be
\label{eq:nonren2}
V(\phi_j)=\sum_{j=0}^N\left(-\tilde m^2\phi_j^{\dagger}\phi_j+\frac{\lambda}{4}|\phi_j^{\dagger}\phi_j|^2\right)+\sum_{j=0}^{N-2}\left(\epsilon \frac{\phi_j^{\dagger}\phi_{j+1}^{q}\phi_{j+2}^{p}}{\Lambda^{p+q-3}}+h.c\right),
\ee
where $p,q$ are integer charges.
Considering again the pions gives:
\be
V(\pi_j)=\frac{m^2}{2}\sum_{j=0}^{N-2}(\pi_j-q\pi_{j+1}-p \pi_{j+2})^2+\mathcal{O}(\pi^4)
\ee
Similarly to the fractional charges case, we have $U(1)^2$ residual symmetry, and two massless modes. Both massless modes will have 'clockwork' behavior in the sense that the coupling to different sites is suppressed by powers of $p$ and $q$, but with no elegant expression. For example for $N=4$ the massless modes of next-to-nearest neighbors are:
 \bea
 \mathcal{O}_{j0}=\mathcal{N}_0\left\{1,\frac{q}{p+q^2},\frac{1}{p+q^2},0,\frac{1}{p(p+q^2)}\right\}\cr
 \mathcal{O}_{j1}=\mathcal{N}_1\left\{1,\frac{p+q^2}{q(2p+q^2)},\frac{1}{2p+q^2},\frac{1}{q(2p+q^2)},0\right\}
 \eea
 The massive modes, are then dominated by $\sim \epsilon m^2(p^2+q^2)$ rather than $\sim \epsilon m^2 q^2$.
 
To recover similar clockwork behavior we choose $q=\tilde q$ and $p=\tilde q^2$, then the potential will be made of terms $V\supset\left(\pi_j-\tilde q\pi_{j+1}-\tilde q^2 \pi_{j+2}\right)^2$. Let us drop the tildes. 
\be
V(\pi_j)=\frac{m^2}{2}\sum_{j=0}^{N-2}\left(\pi_j-q\pi_{j+1}-q^2 \pi_{j+2}\right)^2+\mathcal{O}(\pi^4)
\ee
The above potential is invariant under $\pi_i\rightarrow \pi_i+\left(\frac{-1\pm \sqrt{5}}{2q}\right)^i$, that includes the golden ratio in the argument.
Diagonalizing the mass matrix will give two massless states. We can always choose an ansatz for a massless eigenvector of the form
\be
\mathcal{O}_{j0}=\left\{a_0, -\frac{a_1}{q},\frac{a_2}{q^2},-\frac{a_3}{q^3}, \cdots, \frac{a_N}{q^N}\right\}\equiv\left\{a_j (-q)^j\right\},\,\,
\text{no summation}
\ee
 This generates a recursion relation between the different $a_i$. In the original CW, this relation is simply $a_j=-a_{j+1}$ and hence, once we fix $a_0=1$ all $a_j$s are fixed and we get $\mathcal{O}_{j0}=\mathcal{N}_0\left\{1,\frac{1}{q}\cdots,\frac{1}{q^N}\right\}$. 
When we consider next-to-nearest neighbors, we have two free parameters $a_0,a_1$, and the recursion relation is a Fibonacci sequence
\be
a_{j+2}=a_j+a_{j+1}\Rightarrow \mathcal{O}_{j0}=\mathcal{N}_0\left\{(-q)^{-j} F_j\right\},\,\,
\text{no summation}
\ee
$F_j$ is the j-th Fibonacci number dictated by $a_0,a_1$. 
For instance $a_0=a_1=1$ gives a slight modification of the original CW mode of
$\mathcal{O}_{j0}=\mathcal{N}_0\left\{1,-\frac{1}{q},\frac{2}{q^2},-\frac{3}{q^3} \cdots\right\}$.
The second massless mode is dictated by a different choice of $a_0,a_1$, that then has to undergo the Gram-Schmidt procedure in order to be orthogonal to the CW mode. 
The two different massless modes do scale with $q^{-1}$ between two successive sites, but there is a difference in the total power of $q$.
For the CW mode the ratio $\mathcal{O}_{00}/\mathcal{O}_{N0}\sim q^N$, while for the second massless mode, one has $\mathcal{O}_{01}/\mathcal{O}_{N1}\sim q^{N-1}$. 
It is easy to understand this behavior by choosing two different zero modes. One corresponding to $a_0=1,a_1=0$ and the other with $a_0=0,a_1=1$. In such case, it is clear that one zero mode will have the highest ratio of $q^N$ and the other with $q^{N-1}$. This behavior persists also after orthogonalization.

As a numerical example, let us consider coupling up to next-to-nearest neighbors and $N=3$. 
The first massless mode is
\be
\mathcal{O}_{j0}=\mathcal{N}_0\left\{1,-\frac{1}{q},\frac{2}{q^2},-\frac{3}{q^3}\right\}
\ee
and the second orthogonal one is:
\bea
\mathcal{O}_{j1}=\mathcal{N}_1\left\{\frac{1}{q}+\frac{2}{q^3}+\frac{6}{q^5},1+\frac{2}{q^4}+\frac{3}{q^6},-\frac{1}{q}+\frac{1}{q^3}+\frac{3}{q^7},\frac{2}{q^2}-\frac{1}{q^4}+\frac{2}{q^6}\right\}.
\eea
The two massive modes are:
\bea
\mathcal{O}_{j2}=\frac{\mathcal{N}_2}{q^2}\left\{1,-(1+q),q(1-q),q^2\right\}\cr\\
\mathcal{O}_{j3}=\frac{\mathcal{N}_3}{q^2}\left\{-1,-(1-q),q(1+q),q^2\right\}
\eea
with masses of $m_{2,3}^2=m^2\left(1 \pm q + q^2 \mp q^3 + q^4\right)$.

As can be seen from the example, the massive modes behavior goes between unity and $q^2$ terms, compared to the original CW, where it consisted of unity and linear terms in $q$, \eqref{rotation}. 
In general, there are $N-1$ massive states that can also be brought to the desired orthonormal form by the Gram-Schmidt procedure. Their mass will be dominated by $q^4$ terms rather than $q^2$ in the original clockwork. The elements in eigenvectors will range between unity and $q^2$.  
The case of the integer charges suffers from the same complexity as the fractional charges of the previous section, which is why we do not write down explicit analytical expressions, but only the leading behavior.
Nevertheless, even without a compact analytic expression for the massive modes, the massless mode and the essence of the CW mechanism exists with the $q^{-j}$ overlap.

The generalization to any number of neighbors interactions is straightforward. Considering $n$ nearest neighbors interactions, the lagrangian will look like:
\be
\label{eq:anynumber}
\mathcal{L}=-f^2\sum_{j=0}^{N}\partial U_j^{\dagger}\partial U_j+\frac{m^2f^2}{2}\left(\frac{f}{\Lambda}\right)^{\sum_{i=0}^n q^i-4}\sum_{j=0}^{N-n}\left(U^{\dagger}_jU_{j+1}^{q_1}\cdots U_{j+n}^{q_n}+h.c\right)
\ee
For $n$ nearest neighbors interactions we will preserve $n$ symmetries and the conserved symmetry group will be some $U(1)^n$. 
This is because we have the shift symmetry as long as
\be
\alpha_j=2\pi \ell+\sum_{i=1}^n q_i\alpha_{j+i}
\ee
Such a generalization allows for various breaking patterns, not necessarily reaching the $U(1)$ of the clockwork type. Once again the $n$ massless modes will be suppressed various powers of the $q_i$s. If we wish to maintain the same clockwork behavior, we require $\alpha_{j+1}/\alpha_j\sim q^{-1}$, which enforces $q_i \sim q^i$. 
For $n$ coupled nearest neighbors, the $n$ massless modes are again spanned by an ansatz $\mathcal{O}_{j0}=\mathcal{N}_0\left\{(-q)^{-j} a^{(n)}_j\right\}$ where $a^{(n)}_j$ is given by the recursion relation $a^{(n)}_j=a^{(n)}_{j-n}-(a^{(n)}_{j-(n-1)}+\cdots a^{(n)}_{j-1})=a^{(n)}_{j-n}-F^{(n-1)}_j$, where $F^{(n)}_j$ being the $n$-step Fibonacci number, i.e. $F^{(n)}_j=\sum_{i=1}^n F^{(n)}_{j-i}$. The $n$ different massless modes are determined by various possibilities of $n$ seed numbers required for the recursion relations, $a^{(n)}_1,a^{(n)}_2...$. 
These massless modes are not orthogonal and are brought to such form using the Gram-Schmidt procedure.  The important point is that the CW overlap of $q^{-j}$ is kept. Furthermore, the 'maximal ratio' between the sites for $n$ massless modes will be $q^N,q^{N-1}\cdots q^{N-n}$.

 Let us discuss the massive modes. The radial modes have the same mass of $m_{r_k}^2\sim f^2$ as long as $\lambda \gg \epsilon\left(\frac{f}{\Lambda}\right)^{\sum_{i=0}^n q^i-4} q^{2n}$. This is trivially satisfied. The axions massive modes will be dominated by\footnote{To be precise the axion masses will always behave as $m^2_{a_k}\sim \epsilon f^2 \left(\frac{f}{\Lambda}\right)^{\sum_{i=0}^n q^i-4}\left[(1+ q^{2n})+\sum_{k=1}^{2n-1}\alpha_k q^k \right]$. So one can always express the mass as $m^2_{a_k}=m^2\left((q-1)^{2n}+\sum_{k=1}^{2n-1}\beta_k q^k\right)$. This property is crucial in proving the mass gap when we consider an extra dimensional interpretation.} $m^2_{a_k}\sim \epsilon f^2 \left(\frac{f}{\Lambda}\right)^{\sum_{i=0}^n q^i-4} q^{2n}$. So, depending on $f,\Lambda,q,$ and $n$ one can push the axial modes towards the radial ones, or make them lighter by increasing $\Lambda$.

Except the case of next to nearest neighbors mentioned in the footnote of $V\sim \left( \phi_i^{\dagger}\phi_{i+1}\phi_{i+2}^2+h.c. \right)$ , \eqref{eq:anynumber} does not come from some renormalizeable simple tree-level theory. 
Lifting \eqref{eq:anynumber} to the UV theory is of the following form:
\be
\label{eq:nonrenany}
V(\phi_j)=\sum_{j=0}^N\left(-\tilde m^2\phi_j^{\dagger}\phi_j+\frac{\lambda}{4}|\phi_j^{\dagger}\phi_j|^2\right)+\epsilon\sum_{j=0}^{N-n}\left(\frac{\phi_j^{\dagger}\prod_{i=1}^n\phi_{j+i}^{q_i}}{\Lambda^{\sum q_i-3}}+h.c\right)
\ee
Once again the CW behavior will force $q_i \sim q^i$.

We are now in the position to add all the interactions with any number of neighbors and preserving a single CW $U(1)$ for integer charges, as we should have done in the first place from the EFT point of view. Unlike the charge assignment of $q^n/n$, here we cannot simply add up previous solutions with any number of neighbors, because generically, not a single $U(1)$ symmetry will be conserved. For a conservation of the $U(1)_{CW}$, we need to add up charges 
in a way that does not violate the symmetry. This can be done if we arrange the beyond nearest neighbors interactions in suitable pairs, for $N+1$ even:
\bea
V &=&\sum_{j=0}^N\left(-\tilde m^2\phi_j^{\dagger}\phi_j+\frac{\lambda}{4}|\phi_j^{\dagger}\phi_j|^2\right) \cr 
&+&\epsilon\left(\sum_{j=0}^{N-1}c_1\phi^{\dagger}_j\phi_{j+1}^{q}+\sum_{j=0}^{N-3}c_3\phi^{\dagger}_j\phi_{j+1}^{q}(\phi_{j+2}^{\dagger}\phi_{j+3}^{q})^{\beta_{33}}+\cdots + c_N\phi^{\dagger}_0\phi_{1}^{q}\left(\phi_{2}^{\dagger}\phi_{3}^{q}\right)^{\beta_{3N}}\cdots\left(\phi_{N-1}^{\dagger}\phi_N^{q}\right)^{\beta_{NN}}+ h.c.\right)\cr
\eea
where the $c_j$ fix the correct mass dimensions with relevant powers of $\Lambda$, and the different $\beta$ correspond to some integer power. If all $\beta_{ij}=1$ then we reproduce the original massless CW mode. The  massive modes will slightly change but still scale as $m^2_{a_k}\sim q^2$, while if we choose $\beta_{ij}$ such that each neighbor has a consecutively 
higher power of $q$, such as $\beta_{33}=q^2$, i.e. $\supset \sum_{j=0}^{N-3}c_3\phi^{\dagger}_j\phi_{j+1}^{q}\phi_{j+2}^{\dagger q^2}\phi_{j+3}^{q^3})$, then the massive modes will scale as $m^2\sim q^{2N}$. Of course there are many intermediate possibilities.
 
Alternatively, we need a theory with different charges. Nearest neighbors CW with different charges is discussed in the appendix. For all possible number of neighbors, the full low energy CW potential will now look like:
\bea
V&=&\frac{m^2 f^2}{2}\sum_{k=1}^N\sum_{j=0}^{N-k}\left(c_kU^{\dagger}_j\Pi_{i=1}^kU_{j+i}^{q_{ki}}+h.c\right)\cr \\
&=&\frac{m^2 f^2}{2}\left(\sum_{j=0}^{N-1}c_1U^{\dagger}_jU_{j+1}^{q_{11}}+\sum_{j=0}^{N-2}c_2U^{\dagger}_jU_{j+1}^{q_{21}}U_{j+2}^{q_{22}}+\cdots + c_NU_0^{\dagger}U_1^{q_{N1}}U_2^{q_{N2}}\cdots U_N^{q_{NN}}+ h.c.\right)\cr
\eea
where $c_k\equiv(f/\Lambda)^{\sum_{i=1}^k q_{ki}-3}$ is the appropriate power of $f/\Lambda$.
This is the most general assignment of charges possible. To keep a single CW massless mode, such that $\pi_j\rightarrow \pi_j+\alpha_j$ we need the simultaneous solution of $N$ equations $\alpha_j=2\pi \ell+\sum_{i=1}^k q_{ki}\alpha_{j+i}$ for $k=1$ to $N$.
These are $N$ equations, but we have $N(N+1)/2$ charges. So it is very easy to fulfill such constraint.
A specific example, with a "single" charge $q$ is as follows:
\bea
V &=&\sum_{j=0}^N\left(-\tilde m^2\phi_j^{\dagger}\phi_j+\frac{\lambda}{4}|\phi_j^{\dagger}\phi_j|^2\right) \cr 
&+&\epsilon\left(\sum_{j=0}^{N-1}c_1\phi^{\dagger}_j\phi_{j+1}^{q}+\sum_{j=0}^{N-2}c_2\phi^{\dagger}_j\phi_{j+1}^{q-1}\phi_{j+2}^{q}+\sum_{j=0}^{N-3}c_3\phi^{\dagger}_j\phi_{j+1}^{q-1}\phi_{j+2}^{q-1}\phi_{j+3}^{q}+\cdots + c_N\phi^{\dagger}_0\phi_{1}^{q-1}\phi_{2}^{q-1}\cdots\phi_N^{q}+ h.c.\right)\cr
\eea
where the $c_j$ fix the correct mass dimensions, with $c_j=\Lambda^{-\{j q-(j-1)-3\}}=\Lambda^{-(j(q-1)-2)}$.

The outcome in all of the above cases, is a potential preserving a single CW $U(1)$, and the massless mode has exactly the same profile as the original CW, $\mathcal{O}_{j0}=\left\{1,1/q,\cdots 1/q^N\right\}$.
So we have constructed a CW model with all possible neighbor interactions.
\section{Extra Dimension Interpretation}
\subsection{Alternating Sign}
Considering the CW theory from a 5D point of view, we cannot reproduce neither the full \eqref{eq:original} nor \eqref{eq:lag1}. We can however get the mass matrix \eqref{pimass} by considering a massless free scalar field in 5D, \cite{Giudice:2016yja}.
We would like to give the generalized clockwork mechanism an extra dimensional interpretation. We therefore introduce alternating signs to the potential and the symmetry.  This can be achieved either by a field redefinition $\phi_i\rightarrow (-1)^i \phi_i$ or by considering a 'negative' charge $q<0$. In both cases this is a redefinition that does not change the physics. However, after redefinition, discretizing the extra dimension, higher derivatives in 5D correspond to couplings beyond nearest neighbors in a straightforward manner. 
The symmetry and potential are now:
\bea
\pi_j&\rightarrow& \pi_j+\frac{1}{(-q)^j} \\
V(\pi_j)&=&\frac{m^2}{2}\sum_{j=0}^{N-2}\left(\pi_j+\frac{q}{2}\pi_{j+1}-\frac{q^2}{2} \pi_{j+2}\right)^2+\mathcal{O}(\pi^4)
\eea
Diagonalizing the mass matrix will give two massless states, one of which is the clockwork with $\mathcal{O}_{j0}=\mathcal{N}_0/(-q)^j$ and $N-1$ massive states, again with the masses distributed with $\Delta m/m_a \sim 1$.
The second massless state is simply taking $q\rightarrow -q$ in \eqref{oj1ortho},  
$\mathcal{O}_{j1}=\frac{\mathcal{N}_1}{(-q)^j}\left[(-2)^j-\frac{\sum_{i=0}^N\left(-2q^{-2}\right)^i}{\sum_{i=0}^Nq^{-2i}}\right]$.
This can again be generalized to any number of neighbors $n$ yielding:
\be
V(\pi_j)=\sum_{j=0}^{N-n}\left(\pi_j-\sum_{k=1}^n\frac{(-q)^k}{n}\pi_{j+k}\right)^2+\mathcal{O}(\pi^4)
\ee
Lifting back to the $U_j$s is trivial:
\be
\mathcal{L}=\sum_{j=0}^{N}\partial U_j^{\dagger}\partial U_j-\sum_{j=0}^{N-n}\left(U^{\dagger}_j \Pi_{k=1}^n U_{j+k}^{(-q)^k/n}+h.c\right)
\ee
Again, for $n$ nearest neighbors interaction, we have a residual $U(1)^n$ symmetry, and we can further break it with different breaking patterns.
If we wish to maintain only the clockwork shift symmetry, we can again sum all neighbors interactions:
\be
\mathcal{L}=\sum_{j=0}^{N}\partial U_j^{\dagger}\partial U_j-\sum_{n=1}^N\sum_{j=0}^{N-n}\left(U^{\dagger}_j \Pi_{k=1}^n U_{j+k}^{(-q)^k/n}+h.c\right)
\ee

The generalization provided here and in the previous section can be applied in a straightforward manner to gauge bosons and gravitons at the linear level, as was done in \cite{Giudice:2016yja}.

\subsection{Higher Derivatives in the Extra Dimension}
Coupling to next to nearest neighbors is a nonlocal interaction, and in the continuum language where $N\rightarrow \infty$, we expect higher order derivatives. Indeed, considering a scalar in $5D$, coupling to two consecutive neighbors will correspond to a $(\partial^2_y\phi)^2$, and coupling to $n$ neighbors to $(\partial_y^n\phi)^2$, where $y$ is the extra dimension.

To see this, let's rewrite a more general form of the clockwork lagrangian with two nearest neighbors interaction:
\bea
\label{beta}
V(\pi_j)&=&\frac{m^2}{2}\sum_{j=0}^{N-2}\left(\pi_j-\beta_1q\pi_{j+1}+\beta_2q^2 \pi_{j+2}\right)^2+\mathcal{O}(\pi^4)\\
\pi_i&\rightarrow& \pi_i+\frac{\alpha_j}{(-q)^j} \\
&\Rightarrow &\alpha_j=\beta_1 \alpha_{j+1}-\beta_2\alpha_{j+2}
\eea
If the last equation is fulfilled, we have a shift symmetry, similar to clockwork, but with somewhat different transformation laws. Similarly, the massless eigenvector behaves like $(-q)^{-j}$ with $\mathcal{O}(1)$ coefficient. 
  Consider a compact extra dimension $-\pi R\leq y\leq \pi R$, and identifying $-y$ with $y$.
 Starting from a five dimensional free scalar field with higher derivatives in the extra dimension, we get:
 \footnote{
An equivalent result is obtained by considering, 
$\mathcal{S}=2\int d^4x \int_0^{\pi R}dy\sqrt{-g}\left\{\left(-\frac{1}{2}g^{MN}\partial_M\phi\partial_N\tilde \phi\right)-\frac{1}{2\Lambda^2}\left(\sqrt{-g_{yy}}g^{yy}\partial_y^2\tilde\phi\right)^2\right\}$.  In such case one has the "correct" two powers of the inverse metric, but the action is still not Lorentz invariant in the full $5D$ spacetime, and considering a Lorentz invariant action will include higher derivative terms in the $4D$ equations of motion, as well as introducing site dependence to the clockwork parameters. }

\bea
\label{eq:5dans}
ds^2&=&X(|y|)(-dt^2+d\vec{x}^2)+Y(|y|)dy^2\\
\label{eq:5d2d}
\mathcal{S}&=&2\int d^4x \int_0^{\pi R}dy\sqrt{-g}\left\{\left(-\frac{1}{2}g^{MN}\partial_M\phi\partial_N\tilde \phi\right)-\frac{1}{2\Lambda^2}g^{yy}\left(\partial_y^2\tilde\phi\right)^2\right\}
\eea
Dimensional analysis requires that the higher derivative term will be suppressed by some dimensionful parameter, like $\Lambda^{-2}$. Due to the explicit appearance of only $\partial_y^2$ Lorentz invariance is broken in the extra dimension. Writing a Lorentz covariant action, for instance $\mathcal{S}\supset (\Box_5\phi)^2$ will result in higher derivative terms from the $4D$ spacetime part in the equations of motion, as well as more complicated coupling between $4D$ spacetime derivates and the extra dimension. Performing a field redefinition to get canonical kinetic terms we get:
\bea
\mathcal{S}&=&-\int_0^{\pi R}dy\int d^4x\left\{(\partial_{\mu}\phi)^2+\frac{X^2}{Y^{1/2}}\left(\partial_y \frac{\phi}{X^{1/2}Y^{1/4}}\right)^2+\frac{1}{\Lambda^2}\frac{X^2}{Y^{1/2}}\left(\partial_y^2 \frac{\phi}{X^{1/2}Y^{1/4}}\right)^2\right\}\cr
&=&-\int_0^{\pi R}dy\int d^4x\left\{(\partial_{\mu}\phi)^2+\frac{X^2}{Y^{1/2}}\left(\partial_y \frac{\phi}{X^{1/2}Y^{1/4}}\right)^2\right\}\cr
&-&\int_0^{\pi R}dy\int d^4x\frac{X^2}{\Lambda^2 Y^{1/2}}\left[\frac{\phi''}{X^{1/2}Y^{1/4}}-2\phi'\frac{(X^{1/2}Y^{1/4})'}{(X^{1/2}Y^{1/4})^2}+\phi\left(-\frac{(X^{1/2}Y^{1/4})'}{(X^{1/2}Y^{1/4})^2}\right)'\right]^2
\eea
where prime denotes a derivative w.r.t $y$. Let us discretize the extra dimension with lattice spacing $a$, such that $\pi R=Na$ and use the notation $F_j=F(j a)$, where $F=X,Y,\phi$ and $j$ runs from zero to $N$.
The original clockwork scenario comprises of the first terms in the action, provided that we identify:  
\be
m_j^2=\frac{N^2 X_j}{\pi^2R^2Y_j},\quad q_j=\frac{X_j^{1/2}Y_j^{1/4}}{X_{j+1}^{1/2}Y_{j+1}^{1/4}} \label{eq:5dparams}
\ee
Discretizing the new term gives:
\be
\mathcal{S}\supset -\sum_{j=0}^{N-2}\int d^4x\frac{N^4X_j}{\pi^4\Lambda^2R^4Y_j}(2/q_{j}^2-v_{j+2})^2\left\{\frac{\phi_{j+2}}{(2/q_{j}^2-v_{j+2})}-\frac{2/q_{j}}{(2/q_{j}^2-v_{j+2})}\phi_{j+1}+\phi_j\right\}^2
\ee
where $v_{j+2}\equiv \frac{X_{j+2}^{1/2}Y_{j+2}^{1/4}}{X_{j}^{1/2}Y_{j}^{1/4}}$. Requiring the deconstruction to be independent of the specific site, enforces $v_{j+2}=q_j^{-2}$, thence:
\be
\mathcal{S}\supset -\int d^4x\sum_{j=0}^{N-2}\frac{N^4X_j}{\pi^4\Lambda^2R^4Y_jq_{j}^4}\left\{q_{j}^2\phi_{j+2}-2q_{j}\phi_{j+1}+\phi_j\right\}^2
\ee 
So to recover \eqref{beta}, we demand $\beta_1=2,\,\beta_2=1$.
The "mass" parameter here is different than the original clockwork scenario, $(M^{(2)})^2_j=m_j^2\left(1+\frac{N^2}{\pi^2R^2\Lambda^2}\right)$ but it is just an overall shift. The massless mode remains massless. Notice that no new charges have been introduced, and therefore the solutions for $X,Y$, giving site independent charges $q$ and masses $m$ for the nearest neighbors interaction are similar to the original proposal \cite{Giudice:2016yja} \footnote{In the continuum, this corresponds to $X(|y|)=Y(|y|)=e^{\frac{-4k |y|}{3}}$.}:
\be
\label{sol_dis}
X_j\propto Y_j\propto e^{\frac{-4k\pi R j}{3N}},\quad q=e^{k\pi R/N},\quad m^2=\frac{N^2}{\pi^2R^2}
\ee
In the original CW, plugging \eqref{sol_dis} into \eqref{massa}, we get that we can take the continuum limit of $N\rightarrow \infty$:
\bea
m^2_{a_j}=m^2\left(1+q^2-2 q \cos \frac{j \pi}{N+1}\right), \quad j=1,\cdots N\cr
m^2_{a_j}\simeq \frac{N^2}{\pi^2R^2}\left((q-1)^2+q\left(\frac{j \pi}{N+1}\right)^2+\mathcal{O}(N^{-3})\right)\simeq k^2+\frac{j^2}{R^2}+\mathcal{O}(N^{-1})
\eea

The next to nearest neighbors interaction gets a leading mass term of the form:
\be
M^2=\frac{N^4}{\pi^4R^4\Lambda^2}.
\ee
So we need to check whether the continuum limit works as well. Both with integer and non-integrer powers, we have shown that the axion masses are dominated by $m^2_{a_k}\sim \epsilon f^2 \left(\frac{f}{\Lambda}\right)^{\sum_{i=0}^n q^i-4}(1+ q^{2n})=m^2\left((q-1)^{2n}+\sum_{k=1}^{2n-1}\beta_k q^k\right)$, where $n$ is the number of nearest neighbors. In the case at hand we discuss next to nearest neighbors so $n=2$. 
The leading behavior will be 
\be
m^2_{a_j}\sim M^2(q-1)^4\sim \frac{N^4}{\pi^4R^4\Lambda^2}\left(\frac{k^4\pi^4R^4}{N^4}\right)\sim \frac{k^4}{\Lambda^2}+\cdots
\ee
giving a finite result. 
The mass gap, is determined by the clockwork spring $k$ and the energy scale $\Lambda$. A general formula for the $\beta_k$ coefficients is too complicated to write down, but given the power of $q$, it will be sub-leading. Based on dimensional grounds and numerical examples, we expect it to contribute $m^2_{a_j}\sim \frac{k^4}{\Lambda^2}+c\frac{k^2 j^2}{\Lambda^2 R^2}\cdots$ where $c$ is some $\mathcal{O}(1)$ coefficient.

One may wonder whether other interesting solutions to \eqref{eq:5d2d}, exist rather than trying to reproduce the original clockwork.

%
\section{O(N) Clockwork}
The clockwork mechanism can be implemented for multiple copies of $O(N)$ models and within a single $O(N)$ model as well. Consider $M+1$ copies of O(N) models:
\be
\mathcal{L}=\sum_{j=0}^M -\frac{1}{2}(\partial_{\mu}\vec\phi)^2+\epsilon\sum_{j=0}^{M-1}\left[\frac{\tilde \mu^2}{2}(\vec \phi_j+q\vec \phi_{j+1})^2+\frac{\tilde g}{4N}(\vec \phi_j+q\vec \phi_{j+1})^4\right]
\ee
For $\epsilon\rightarrow 0$ we have a global $O(N)^{M+1}$ symmetry as well as shift symmetry for every vector $(\mathbb{R}^N)^{M+1}$. Turning on $\epsilon$ breaks the symmetry down to a single $O(N)$, since all vectors have to be rotated by the same orthogonal matrix. However, there is still a shift symmetry of $\vec \phi_j\rightarrow \vec \phi_j+\frac{\vec{c}}{(-q)^{j}}$, where $\vec{c}$ is a constant vector. So the full symmetry group is now $O(N)\ltimes\mathbb{R}^N$ The eigenvector corresponding to the conserved $O(N)$ remains massless, and this vector will have the same $q^{-j}$ suppression in overlap with the different $\vec \phi_j$.

\subsection{Linear Sigma Model}
The original clockwork discusses the breaking of $U(1)^N$ after spontaneous symmetry breaking.
We now generalize it to $M+1$ copies of O(N) models in the broken phase.
Consider $M+1$ copies of O(N) models:
\be
\label{eq:ONoriginal}
\mathcal{L}=\sum_{j=0}^M \left(-\frac{1}{2}\left(\partial_{\mu}\vec\phi\right)^2-\frac{\mu^2}{2}\left(\vec \phi_j\right)^2+\frac{\lambda}{4}\left(\vec{\phi_j}^2\right)^2\right)+\epsilon\sum_{j=0}^{M-1}\left[\frac{\tilde \mu^2}{2}\left(\vec \phi_j-q\vec \phi_{j+1}\right)^2+\frac{\tilde g}{4N}\left(\vec \phi_j-q\vec \phi_{j+1}\right)^4\right]
\ee
Notice that here, the shift symmetry from the previous paragraph is gone.
We use the conventional parametrization where the Nth field gets a vev $v_i$:
\be
\vec{\phi_i}=\left(\pi_{ik},v_i+\sigma_i\right),\quad v_i=\frac{\mu}{\sqrt{\lambda}}
\ee
For simplicity, we took all the vevs to be the same. It is a trivial generalization to consider for each model a different $\mu_i,\lambda_i$ such that the vevs will be different.
In this case, when $\epsilon \rightarrow 0$ we have $M+1$ copies of O(N) models in their broken phase, such that there is 
a global $O(N-1)^{M+1}$ symmetry, and there are $(M+1)\times (N-1)$ massless goldstone bosons. These are the $\pi_{ik}$.
Turning on the clockwork terms induces a coupling that will break the symmetry explicitly to a single $O(N-1)$.
However, the analysis of the massless modes and their localization is not as trivial. The CW term induces a nonzero vev for the $\sigma_i$ fields and ruins the localization. To avoid this, we need a vanishing vev for these fields also after turning on the CW term.
This is doable. 
Consider the following potential:
\be
V=\sum_{j=0}^M \left(-\frac{\mu_0^2}{2q^{2j}}\left(\vec \phi_j\right)^2+\frac{\lambda}{4}\left(\vec{\phi_j}^2\right)^2\right)+\epsilon\sum_{j=0}^{M-1}\left[\frac{\tilde \mu^2}{2}\left(\vec \phi_j-q\vec \phi_{j+1}\right)^2+\frac{\tilde g}{4N}\left(\vec \phi_j-q\vec \phi_{j+1}\right)^4\right]
\ee
Notice that here the quadratic term is different for each $O(N)$ vector, with $\mu_j=\frac{\mu_0}{q^j}$\footnote{We could have constructed the desired behavior by having $\lambda_j=\lambda_0q_j^2$ or any other combination that gives $v_j=v_0/q^j$.} Obviously this is less robust than \eqref{eq:ONoriginal}, where in principle every $O(N)$ vector could have any quadratic term.
Using again the conventional parameterization 
\be
\vec{\phi_i}=\left(\pi_{ik},v_i+\sigma_i\right),\quad v_i=\frac{\mu_i}{\sqrt{\lambda}}=\frac{\mu_0}{q^i\sqrt{\lambda}},
\ee
 ensures $<\sigma_j>=0$ even in the presence of the CW term. 
  To demonstrate this, let us reinsert the conventional parameterization into the potential:
 \bea
 V&=&\sum_{j=0}^M \left[-\frac{\mu_0^2}{2q^{2j}}\left(\pi_{ij}\pi_{ij}+(v_j+\sigma_j)^2\right)+\frac{\lambda}{4}\left(2(v_j+\sigma_j)^2\pi_{ij}\pi_{ij}+(v_j+\sigma_j)^4+{\mathcal O}(\pi^4)\right)\right]\cr
 &+&\epsilon\sum_{j=0}^{M-1}\Bigg[\frac{\tilde \mu^2}{2}\left\{\left(\pi_{ij}-q\pi_{ij+1}\right)^2+(v_j+\sigma_j-q v_{j+1} -q \sigma_{j+1})^2\right\}\cr
 &+&\frac{\tilde g}{4N}\left\{2\left(\pi_{ij}-q\pi_{ij+1}\right)^2(v_j+\sigma_j-q v_{j+1} -q \sigma_{j+1})^2+(v_j+\sigma_j-q v_{j+1} -q \sigma_{j+1})^4+{\mathcal O}(\pi^4)\right\}\Bigg]\cr
 \eea
 Since $v_j=v_0/q^j$ all the vev terms in the CW term vanish and we are left with:
 \bea
 V&=&\sum_{j=0}^M \left[-\frac{\mu_0^2}{2q^{2j}}\left(\pi_{ij}\pi_{ij}+(v_j+\sigma_j)^2\right)+\frac{\lambda}{4}\left(2(v_j+\sigma_j)^2\pi_{ij}\pi_{ij}+(v_j+\sigma_j)^4+{\mathcal O}(\pi^4)\right)\right]\cr
 &+&\epsilon\sum_{j=0}^{M-1}\Bigg[\frac{\tilde \mu^2}{2}\left\{\left(\pi_{ij}-q\pi_{ij+1}\right)^2+(\sigma_j -q \sigma_{j+1})^2\right\}\cr
 &+&\frac{\tilde g}{4N}\left\{2\left(\pi_{ij}-q\pi_{ij+1}\right)^2(\sigma_j -q \sigma_{j+1})^2+(\sigma_j -q \sigma_{j+1})^4+{\mathcal O}(\pi^4)\right\}\Bigg]\cr
 \eea
 The CW term is positive semi-definite and obviously $<\pi_{ij}>=<\sigma_j>=0, \, \forall i,j$ is the minimum of the CW term\footnote{It is also clear that $<\pi_{ij}>=<\sigma_j>=0, \, \forall i,j$ is at least an extremum regardless of the signs of the different terms in the potnetial.} and therefore the full potential.
There are no new tadpole contributions coming from the CW term. 
This is because we can write the quadratic CW term (of both $\pi$s and $\sigma$s) as follows:
\be
\begin{pmatrix}
\vec{\phi_0} & \cdots & \vec{\phi_M}
\end{pmatrix}^T
\times 
\begin{pmatrix}
1 & -q & 0 &\cdots &0\\
-q &1+q^2 & -q & 0 &\cdots  \\
0 &-q &1+q^2 &-q &0 \, \cdots \\
\cdots & \cdots & \cdots & \cdots & \cdots\\
\cdots & \cdots & -q & 1+q^2 & -q\\
 \cdots & \cdots & \cdots & -q& q^2
 \end{pmatrix}
 \begin{pmatrix}
\vec{\phi_0} & \cdots & \vec{\phi_M}
\end{pmatrix}
 \ee
 where each entry in the matrix corresponds to a vector of length $N$.  The $\sigma$ are all massive fields since they have an additional contribution from the standard $O(N)$ model terms. Finally, we have a single $O(N-1)$ and a single linear combination of 
$\pi_{ik}$ massless, again with overlap suppressed with the standard $q^{-j}$. The eigenvalues of such a matrix will be the same as the clockwork, with a massless $O(N-1)$ vector.
As a numerical example, consider the case of $N=4$, $M=2$.
We shall have a block diagonal mass matrix for the $\pi$s, where each block is:
\be
\begin{pmatrix}
1 & -q & 0  \\
-q &1+q^2 & -q    \\
0 &-q &q^2    \\
 \end{pmatrix}
 \ee
giving us an $O(3)$ CW massless vector.

 Establishing the CW behavior for a global $O(N)$ symmetry, one can still have exponential seclusion of the massless mode, unlike the gauge symmetry result reported in \cite{Craig:2017cda}. Considering, for instance,
\be
\mathcal{L}\supset -\frac{1}{4g^2}F_{\mu \nu}F^{\mu \nu}+\frac{(\vec \phi_N)^2}{8\pi f^2}F_{\mu \nu}\tilde F^{\mu \nu}
\ee
the massless mode coupling will behave as
\be
\mathcal{L}\supset \frac{a_0^2}{8\pi f_0^2}F_{\mu \nu}\tilde F^{\mu \nu}, \quad f_0=q^Nf
\ee

\section{Clockwork SUGRA}
\subsection{Canonical K\"ahler Potential}
\label{trivialsugra}
Clockwork mechanism in global SUSY has been suggested in \cite{Kaplan:2015fuy}.
Considering $3(N+1)$ chiral superfields $S_j, \Phi_j,\tilde \Phi_j$. For $q=2$ one can write down the renormalizeable superpotential:
\be
\label{eq:sugra}
W=\sum_{j=0}^N\lambda S_j\left(\Phi_j \tilde \Phi_j-v^2\right)+\epsilon\sum_{j=0}^{N-1}\left(\Phi_j\tilde \Phi_{j+1}^2+\tilde \Phi_j \Phi_{j+1}^2\right)
\ee
Taking $\epsilon \rightarrow0$ reveals a $U(1)^{N+1}$ global symmetry. Turning on $\epsilon$ breaks these symmetries into a single $U(1)$  with hierarchical charges with $S_j$ being neutral, $\Phi_j$ a charge of $2^{-j}$ and $\tilde \Phi_j$ with charge $-(2)^{-j}$. It is interesting to note, that taking $\lambda\rightarrow 0$, produces \textit{two} $U(1)$ symmetries, unlike the non-supersymmetric case, where there is only a single $U(1)$ \cite{Farina:2016tgd}. The requirement for a SUSY minimum $W_i=0$ for all chiral superfields gives the vev $\Phi_j \tilde \Phi_j=v^2$ as well as non-zero $S_j$, see below. SUSY is conserved.
We then have $2(N+1)$ massive chiral superfields and $N+1$ massless ones. 
 The low energy theory below the scale $\lambda v$ can then be parameterized as 
$\Phi_j=ve^{\Pi_j/v},\tilde \Phi_j=ve^{-\Pi_j/v}$, yielding:
\be
W=2\epsilon v^3\sum_{j=0}^{N-1}\cosh\left(\frac{\Pi_j-2\Pi_{j+1}}{v}\right)
\ee
Notice, that we still have a Minkowski SUSY minimum for the low energy theory as well. 

However, turning on gravity, we shall see that this construction is insufficient. We encounter the well known problem of getting AdS SUSY vacuum. To get a viable phenomenology, we must uplift this minimum into a SUSY breaking Minkowski or dS minima. As such, adding a constant term in the superpotential $W$ to uplift the vacuum and building a CW model on top of that is simply a test of the uplift term. We therefore seek alternative constructions of the clockwork within SUGRA.

The simplest generalization to SUGRA is straightforward.  Throughout this section, we consider the Planck mass to be unity. Consider a canonical K\"ahler potential $K=\sum_{j=0}^N |\Phi_j|^2+|\tilde \Phi_{j}|^2+|S_j|^2$.
The K\"ahler potential $K$ is invariant under the same $U(1)^{N+1}$ as the superpotential for $\epsilon \rightarrow 0$ case, as well as additional $U(1)$ for each chiral superfield, to a total of $U(1)^{3(N+1)}$. 
The F-term scalar potential reads:
\be
V  =  e^{K}(D_{i}WD_{\bar{j}}\overline{W}K^{i\bar{j}}-3|W|^{2})
\ee
The requirement for a supersymmetric minimum is now changed to $D_iW=0$ for all chiral superfields (no summation):
\bea
D_{S_j}W&=&\lambda \left(\Phi_j \tilde \Phi_j-v^2\right)+\overline{S}_{\bar j}W \label{dskahler}\\
D_{\Phi_j}W&=&\lambda S_j \tilde \Phi_j+\epsilon \left(\tilde \Phi_{j+1}^2+2\Phi_j\tilde \Phi_{j-1}\right)+\overline{\Phi}_{\bar j}W \label{dphikahler}
\eea
$K,W$ are invariant on interchanging $\tilde \Phi_j$ and $\Phi_j$, so we just consider the fields without the tilde.
Considering the case of $\epsilon\rightarrow0$, it is clear that the only supersymmetric minimum is that of global SUSY with $W_i=W=0$ at the minimum and $<S_j>=0,\, <\Phi_j \tilde{ \Phi}_j>=v^2$.
We now need to check whether the different terms in the scalar potential are still invariant under the clockwork $U(1)$. Since the K\"ahler potential $K$ and the superpotential $W$ are invariant, and the K\"ahler metric is a unit metric, we just need to check the K\"ahler derivative. The $S_j$ K\"ahler derivative is invariant. Regarding the $\Phi_j$:
\be
\Phi_i\rightarrow e^{i\alpha_i}\Phi_i,\quad
\tilde \Phi_i\rightarrow e^{-i \alpha_i}\tilde \Phi_i,\quad
\alpha_i=2\alpha_{i+1},\quad
D_{\Phi_i}W \rightarrow e^{-i\alpha_i}D_{\Phi_i}W \Rightarrow |D_iW|^2\rightarrow |D_iW|^2
\ee
Thus, the entire scalar potential and kinetic terms are invariant under the clockwork symmetry.
Let us now consider the $\epsilon \neq 0$ case in \eqref{eq:sugra}. First, let us look for a global SUSY minimum, i.e. $W=W_i=0$. 
From \eqref{dskahler} we get $<\Phi_j \tilde{ \Phi}_j>=v^2$. Substituting this vev into \eqref{dphikahler} shows that vanishing $S_j$ is not a solution anymore. Rather, $<S_0>=-\epsilon v/\lambda, \, <S_N>=-2\epsilon v/\lambda,\, <S_j>=-3\epsilon v/\lambda,\, \forall j\neq0,N$. Substituting these vevs into $W$ we get $W\neq 0$, so we have a contradiction. 
However, this can readily be fixed. 
Consider adding a constant term to the superpotential:
\be
W=w_0+\sum_{j=0}^N\lambda S_j\left(\Phi_j \tilde \Phi_j-v^2\right)+\epsilon\sum_{j=0}^{N-1}\left(\Phi_j\tilde \Phi_{j+1}^2+\tilde \Phi_j \Phi_{j+1}^2\right)
\ee
The purpose of the additional constant $w_0$ is to make sure that $W=0$ at the supersymmetric minimum, thus ensuring a global SUSY solution.
Demanding $W_i=W=0$ for the SUSY minimum for all chiral superfields gives $\Phi_i=\tilde \Phi_i=v$. The $S_i$ now receive non-zero vevs:
\be
S_0=-\frac{\epsilon v}{\lambda},\quad S_N=-\frac{2 \epsilon v}{\lambda},\quad S_i=-\frac{3 \epsilon v}{\lambda} \quad \forall i\neq 0,N.
\ee
This fixes the constant $w_0$ to be $w_0=-2N\epsilon v^3$.
Thus, we have trivially generalized the CW scenario to SUGRA at the price of adding an arbitrary tuned constant $w_0$ to the superpotential $W$.
This constant $w_0$ is parametrically the same as the CW term.

Pursuing such phenomenology is viable, but hinges on the exact value of $w_0$, whatever its origin is \cite{Brustein:2004xn,BenDayan:2008dv}. It makes sense to consider alternatives that do not rely on $w_0$, which we turn to next. 

\subsection{Shift Symmetric K\"ahler Potential}
\label{shiftK}
 Shift symmetries in the K\"ahler potential are abundant in SUGRA constructions, especially in the inflationary literature, \cite{Kawasaki:2000yn,Kallosh:2010xz,BenDayan:2010yz}.
A possible construction is by using a so-called "stabilizer" field, eloquently explained in \cite{Kallosh:2010xz}.
Consider a superpotential and K\"ahler potential of the form:
\be
W=Sf(\Phi_i),\quad K=S\bar{S}+\sum_i\frac{1}{2}\left(\Phi_i+\bar{\Phi}_{\bar{i}}
\right)^2
\ee
In such case, we have a Minkowski SUSY vacuum at  , the kinetic terms of the fields are canonical and the potential at $S=\Re(\Phi_i)=0$ is simply:
\be
V=f^2(\Im \Phi_i)
\ee
Hence, by choosing the function $f$ to have only arguments of the form $f=f(\Phi_i-q\Phi_{i+1})$, we can have potentials that manifestly have the shift symmetry of $\Phi_i \rightarrow \Phi_i+c/q^{i}$.

The structure of the vacuum in this case is rather generic. Assuming a Minkowski vacuum, $V|_0=0\Rightarrow f|_0=0$.
In such a case, at the extremum $V_i=2ff_i=0$, and thus the mass matrix at the vacuum becomes:
\be
V_{ij}|_0=2(ff_{ij}+f_if_j)|_0=2f_if_j|_0
\ee
In such case we do not have one flat direction and a single massless modes, but the usual CW shift symmetry and $N$ massless modes! 
There is no hierarchy generated between a single massless mode and $N$ massive states.
We can uplift these $N$ massless modes by explicitly breaking SUSY using another CW coupling:
\be
V(\Im \Phi_i)=f^2(\Im \Phi_i-q \Im \Phi_{i+1})+\epsilon(\Im \Phi_i-p\Im \Phi_{i+1})^2
\ee
where $p$ is the new CW charge, and $\epsilon \ll 1$ is a small SUSY breaking parameter giving us finally, an embedding of CW in SUGRA.
This is of course a fine-tuned construction.

Alternatively, we can have a small CC, such that $f|_0=W_0$.\footnote{
Notice that here, $W_0$ has mass dimension two, while in the previous subsection, $w_0$ had mass dimension three.} Obviously here, we do not have a SUSY Minkowski minimum (with $S=f(\Phi_i)_0=0$), but rather a dS SUSY breaking one, (with $S=0,f(\Phi_i)\neq 0$)
In such case:
\bea
V|_0&=&W^2_0\cr
V_i|_0&=&2ff_i|_0=0\Rightarrow f_i|_0=0\cr
V_{ij}|_0&=&2W_0f_{ij}|_0 \label{Hessian}
\eea

Arranging $f_{ij}$ to have positive semi-definite mass matrix is easy, for instance, expanding around the minimum, we can simply take $f=W_0+\sum \alpha(\Phi_i-q\Phi_{i+1})^2$, so we have a massless CW mode and $N$ massive states, according to \eqref{massa}
\be
m^2_{a_0}=0, \quad m^2_{a_k}=2W_0\alpha\left(1+q^2-2 q \cos \frac{k \pi}{N+1}\right), \quad k=1,\cdots N
\ee
However, as we can see the mass is related to the CC $\Lambda^4 \sim W_0^2$.
So the fundamental scale of the CW sector is parametrically connected to the CC.
 
 \subsection{Shift Symmetric Superpotential}
Whereas in \ref{trivialsugra} $w_0$ was a parameter ensuring a Minkowski SUSY mimimum, in the shift symmetric K\"ahler potential case, it resulted in an actual CC.  Here we try a different approach and for $N+1$ chiral superfields consider a manifestly symmetric superpotential of the form:
 \be
 W=\sum_{i=0}^{N}m(\Phi_i-q\Phi_i)^2
 \ee
 The superpotential is invariant under the shift symmetry $\Phi_i\rightarrow \Phi_i+\alpha_i/q^i$.
A Minkowski vacua with $W=W_i=0$ at the minimum exist for vevs $\langle \Phi_i\rangle=q\langle \Phi_{i+1}\rangle$, with some $\langle \Phi_0\rangle \equiv v$ \footnote{Any $\langle \Phi_0\rangle \equiv v$ including a vanishing $v$ is a solution,  so we have a flat direction with degenerate Minkowski vacua}.  
 For any K\"ahler potential, we shall have the following potential and derivatives \cite{BenDayan:2008dv}:
 \begin{eqnarray}
V & = & e^{K}(D_{i}WD_{\bar{j}}\overline{W}K^{i\bar{j}}-3|W|^{2})\label{eq:P}\\
\partial_{k}V & = & e^{K}(D_{k}D_{i}WD_{\bar{j}}\overline{W}K^{i\bar{j}}-2D_{k}W\overline{W})\label{eq:dP}\\
\nabla_{l}\partial_{k}V & = & e^{K}(D_{l}D_{k}D_{i}WD_{\bar{j}}\overline{W}K^{i\bar{j}}-D_{l}D_{k}W\overline{W})\label{eq:ddP}\\
\nabla_{\bar{l}}\partial_{k}V & = &
e^{K}(-R_{k\bar{l}i\bar{m}}D_{n}WD_{\bar{j}}\overline{W}K^{i\bar{j}}K^{n\bar{m}}+K_{k\bar{l}}D_{i}WD_{\bar{j}}\overline{W}K^{i\bar{j}}-D_{k}WD_{\bar{l}}\overline{W}\nonumber
\\
 &  & +D_{k}D_{i}WD_{\bar{l}}D_{\bar{j}}\overline{W}K^{i\bar{j}}-2K_{k\bar{l}}W\overline{W}).
 \label{eq:ddbarP}
 \end{eqnarray}
In the above $\partial_{i}$ denotes differentiation with respect
to a chiral scalar $\phi^{i}$, $K_{i}=\partial_{i}K$ etc. and
\begin{eqnarray}
D_{i}X_{j}&=&\nabla_{i}X_{j}+K_{i}X_{j} \cr
\nabla_{i}X_{j}&=&\partial_{i}X_{j}-\Gamma_{ij}^{k}X_{k} \cr
\Gamma_{ij}^{k}&=&K^{k\bar{l}}\partial_{i}K_{j\bar{l}} \\
R_{i\bar{j}k\bar{l}}&=&K_{m\bar{l}}\partial_{\bar{j}}\Gamma_{ik}^{m}. \nonumber
\label{eq:Kahlergeom}
\end{eqnarray}

Evaluating these quantities at the Minkowski SUSY vacuum for \textit{any} K\"ahler potential gives:
\be
V=\partial_{k}V=\nabla_{l}\partial_{k}V =W=W_i=0
\ee
and the only nonzero term is:
\be
\nabla_{\bar{l}}\partial_{k}V|_{0}=e^{K}K^{i \bar{j}}W_{ik}\overline W_{\bar{j}\bar{l}}|_{0}
\ee
where $W_{ij}$ is a matrix exactly of the form \eqref{pimass}, with $m$ as the mass parameter instead of $m^2$. Hence, for canonical K\"ahler, the mass matrix for the scalars will be:
\bea
\label{mijbar}
m_{i \bar{j}}^2=e^{K|_0} |m|^2
\left|\begin{pmatrix}
1 & -q & 0 & \cdots &  & 0 \cr
-q & 1+q^2 & -q & \cdots &  & 0 \cr
0 & -q & 1+q^2 & \cdots & & 0 \cr
\vdots & \vdots & \vdots & \ddots & &\vdots \cr
 & & & & 1+q^2 & -q \cr
 0 & 0 & 0 &\cdots & -q & q^2
\end{pmatrix}\right|^2 \, .
 \eea
 with guaranteed clockworking, a single massless complex superfield $m^2_{a_0}=0$ and $N$ parametericaly heavy ones. The masses will be the square of the original CW masses for real fields, $ m^2_{a_k}=e^{K|_0} |m|^2\left(1+q^2-2 q \cos \frac{k \pi}{N+1}\right)^2$.
The main difference compared to the previous section, was the existence of a scale $m$ instead of the stabilizer field $S$. 

 For other K\"ahler manifolds, one has to look more carefully at the mass matrix.
 In the most general case, the canonically normalized mass matrix is \cite{BenDayan:2008dv}:
 
 \bea
\mathcal{M}=\left(\begin{array}{cc} \vspace{.05in}
K^{i \overline m} N_{\overline m j} & K^{{i} \overline m} N_{\overline{m} \overline{j}} \\
K^{\overline{i}  m} N_{m  j } & K^{\overline i  m} N_{  m  \overline j}
\end{array}\right)
\eea
with
\bea
N_{i\overline j}&=&\nabla_i \partial_{\overline j} V\cr
N_{i  j}&=&\nabla_i\partial_{ j}V-\Gamma_{i j }^{\ \  k}\partial_{{k}}V.
\eea
As before $N_{ij}=0$ at the vacuum, while $N_{i \overline{j}}=m_{i \bar{j}}^2$ from \eqref{mijbar}. Thus, for canonical K\"ahler, we get the clockwork. For a diagonal K\"ahler each mode $\Phi_i$ will get multiplied by the corresponding inverse K\"ahler term $K^{i \bar{i}}|_{0}$ (no summation), and the massless mode will have different weighting of each field.
Finally for a general K\"ahler potential, we are still guaranteed a single complex massless mode, since the K\"ahler metric $K_{i \bar{j}}$ is invertible, but whether the exponential suppression exists and the corresponding overlap of each site depends on the specific K\"ahler potential considered.

 To summarize, embedding the CW mechanism in SUGRA is a problematic issue.  According to \ref{trivialsugra},\ref{shiftK} we either a. must add some constant as large as the CW term to "fix" the CC, or b. get many flat directions and no clockworking. One has to further introduce explicit SUSY breaking CW to generate the desired CW hierarchy. 
 
 
 To avoid such conclusion, we must manifestly build the CW symmetry into the superpotential $W$ along with a scale $m$ that does not come from a vev of another chiral superfield. In such case, we get the correct CW mass matrix for the case of canonical K\"ahler potential.
 
 \subsection{SUSY Breaking}
 As everyone knows, SUSY is broken in nature, so the analysis will not be complete without incorporating a SUSY breaking mechanism. Specifically, it is important to analyze the connection between SUSY breaking and the CW symmetry. Can we have one without the other? 
 An immediate SUSY breaking mechanism, is again provided by $w_0$ of section $5.1$. If we do not tune the constant to be $|w_0|=2N\epsilon v^3$, but somewhat smaller, then we will not get an exact cancellation of the CC, and SUSY will be broken exactly by this residual. 

\subsubsection{SUSY breaking with CW symmetry conservation using a spurion}
A more interesting case is reconsidering the shift symmetric superpotential $W=\sum_{i=0}^{N}m(\Phi_i-q\Phi_{i+1})^2$. SUSY breaking can be achieved in a straightforward way by spurion analysis with a canonical K\"ahler. 
 Consider SUSY breaking, while the CW symmetry is left untouched:
\be
K=\sum_{i=0}^{N}\Phi_i\bar \Phi_i+S \bar S, \quad W=\sum_{i=0}^{N}m(\Phi_i-q\Phi_{i+1})^2 +\delta m^2 S, 
 \ee
 where $\delta m \ll m$. 
 All the $\Phi_i$ fields still fulfill $D_i W=0$ if $\langle \Phi_i\rangle=q\langle \Phi_{i+1}\rangle$, for any $\langle \Phi_0\rangle \equiv v$.
However, due to the appearance of the spurion, an actual minimum occurs only if all the vevs of all fields including $S$ vanish.
Then SUSY is spontaneously broken, with $W|_{vac.}=W_{\Phi_i}|_{vac.}=0$ and $W_S=\delta m^2$. 
In the vacuum we shall have  
\be
V=e^K|W_S|^2 \quad \textit{and} \quad V_{vac.}=|\delta m|^4>0
\ee

It is obvious that the CW shift symmetry of $W$ is unharmed. 
 The SUSY breaking scale is controlled by $\delta m$ that is much smaller than the rest of the scales in the problem. From this point, SUGRA model building with collider signature analysis can ensue in a standard way. For example soft SUSY breaking terms can break the CW symmetry and lead to viable models.
 
\subsubsection{SUSY breaking with CW breaking}
Let us now try to induce SUSY breaking using the CW fields themselves and breaking the CW symmetry at the same time. For a viable SUSY breaking with a positive CC, we shall see that the SUSY breaking direction is special and this direction is exactly the massless eigenmode of the clockwork!

Lets assume a quadratic K\"ahler, i.e. 
$K=\sum_{i=0}^{N}\Phi_i\bar \Phi_i$ or $K=\sum_{i=0}^{N}\frac{(\Phi_i+\bar \Phi_i)^2}{2}$ or any mixture of such potentials, as long as $K_{i \bar j}=\delta_{i \bar j}$, with the following superpotential, written around the desired SUSY breaking minimum
\be
W=\sum_{i=0}^{N}m(\Phi_i-q\Phi_{i+1})^2 +\sum_{i=0}^{N} c_i\Phi_i
\ee
For $c_i\rightarrow 0$, both SUSY and the CW shift symmetry of the superpotential are conserved. For $c_i\neq 0$ the CW shift symmetry is broken. Let us analyze the SUSY breaking.
Obviously, at $\Phi_i=0$, we shall have $W|_0=0$ and $W_i|_0=c_i$. Thus $D_iW|_0=c_i$ and all fields collectively participate in SUSY breaking. 
To find a minimum, we refer again to \eqref{eq:dP}:
 \be
 \partial_{k}V =  e^{K}(D_{k}D_{i}WD_{\bar{j}}\overline{W}K^{i\bar{j}}-2D_{k}W\overline{W})=0
 \ee
The second term always vanishes since $W|_0=0$, so we are left with:
\be
0=D_{k}D_{i}WD_{\bar{j}}\overline{W}K^{i\bar{j}}=D_kD_iWc_i=(W_{ki}+K_{ki}W+K_iW_k)c_i=W_{ki} c_i
\ee

Hence, we got an eigenvalue problem with eigenvalue zero, of the matrix $W_{ij}$ which is exactly \eqref{pimass} (only with $m$ instead of $m^2$):
\be
\label{eq:ev}
(M^2_\pi)_{ij}c_j=0
\ee
The solution of the zero eigenvalue is exactly the CW vector, $c_i=v^2/q^i$ where $v$ is some constant.
Thus, we found that for a canonical K\"ahler with the CW superpotential, there is a unique SUSY breaking direction, that is the CW direction:
\be
W=\sum_{i=0}^{N}m(\Phi_i-q\Phi_{i+1})^2 +\sum_{i=0}^{N} \frac{v^2}{q^i}\Phi_i
\ee
Any single or several $c_i\Phi_i$ will not yield a good SUSY breaking minimum, as it will not fulfill \eqref{eq:dP}, or \eqref{eq:ev}. i.e. this model has the simplest realization of SUSY and CW symmetry being broken at a single stroke and from within the CW gears, rather than creating a hierarchy of symmetry breaking, or adding fields 'external' to the CW lagrangian. This is a unique breaking pattern, that limits model building possibilities.
In such case, it is actually easy to write $W$ in the mass basis. Denoting the rotated superfields in the mass basis by $\tilde a_i$, the superpotential is simply:
\be
W=\tilde v^2 \tilde a_0+\sum_{i=1}^{N}m_{i}\tilde a_i^2
\ee
where $m_i$ are the masses written in \eqref{massa}, with $m$ instead of $m^2$ and $\tilde v^2=v^2/\mathcal{N}_0$, using a proper normalization of $\tilde a_0$. 
It is clear that all $\tilde a_{i\neq 0}$ will sit at their supersymmetric minimum at $\tilde a_i=0$, while only the CW mode will break SUSY, as well as the flat direction of $\tilde a_0$ which one has for $v\rightarrow 0$. The minimum of the scalar potential is now at $V|_0=v^4\sum_{i=0}^Nq^{-2i}=\tilde v^4>0$. From here, model building continues in a standard way. 
\subsubsection{SUSY breaking in the mass basis}
Once we move to the mass basis, it is clear how to construct spontaneously broken SUSY using the CW fields, while preserving the CW symmetry. Note that a canonical K\"ahler of the form $K=\sum_{i=0}^{N}\Phi_i\bar \Phi_i$ becomes $K=\sum_{i=0}^{N}\tilde a_i \bar{\tilde a}_i$.
All one has to do is choose one of the $\tilde a_{i\neq 0}$ to be the spurion:
\be
W=\delta m^2(\tilde a_i-c)+\sum_{i=1}^{N}m_{i}\tilde a_i^2.
\ee
For proper choices of $\delta m^2, c$, one has a SUSY breaking solution. As we can see, this requires a tuned constant $w_0\equiv -\delta m^2 c$.
The massless CW mode $\tilde a_0$ still does not appear in $W$, that still possesses a CW shift symmetry. 

To summarize, we see that either SUSY is broken by some spurion outside the CW sector where the CW symmetry is conserved, or the CW mechanism, dictates the SUSY breaking direction. Without $w_0$, the SUSY breaking direction is exactly the original massless CW mode, that plays the role of the spurion, leaving no residual CW symmetry. With $w_0$, SUSY breaking can occur along any of the $\tilde a_i$ directions and the CW shift symmetry is conserved.

\section{Conformal Coupling and Mass Terms in the Extra Dimension}
It is interesting to contemplate the possibility of some generic explicit breaking of the residual $U(1)$ symmetry, beyond the direct coupling to the specific sector we are interested in. If so, we do not only have technical naturalness, but can potentially have a 'natural' UV theory that includes the breaking of the CW symmetry. Such a possibility could come out directly from the extra dimensional interpretation. 
Considering again the extra dimensional picture, \cite{Giudice:2016yja} reproduced the mass matrix of the $4D$ CW and suggested that the continuum is coming from a linear dilaton model. 

 Because the main aim of this paper is to generalize CW theory, it makes sense to consider how additional terms in 5D affect the CW structure. Since, we can at most recover the mass matrix of CW, we  limit ourselves to terms that are at most quadratic in the 5D scalar field. A free massless scalar field in 5D generated the CW mass matrix \eqref{pimass}. We therefore expect the additional quadratic terms to generate an explicit breaking of the CW symmetry as desired. 
We explore the effect of a conformal coupling to gravity, as the most minimal variation, where conformal symmetry, or to be precise, local Weyl symmetry, $g_{MN}\rightarrow e^{2\omega(x)}g_{MN},\, \phi\rightarrow e^{-3/2\omega(x)}$ is maintained. \footnote{Some analysis along these lines with the Randall-Sundrum (RS) metric has been carried out in \cite{Hofmann:2003gk}.} Deconstruction will introduce a length scale, the lattice spacing, and will break the conformal symmetry, and generate a $4D$ mass term. 
We then consider explicitly adding mass terms in the extra dimension. Finally, we deviate from the metric suggested in \cite{Giudice:2016yja}, and consider positively curved 5D manifolds.

The 5D dimensional action of a scalar field conformally coupled to gravity is:
\be
\label{eq:Sconformal}
\mathcal{S}=-2\int_0^{\pi R}dy\int d^4x\sqrt{-g_5}\left[g^{MN}\partial_M\phi\partial_N \phi+\frac{\xi}{2}\phi^2\mathcal{R}_5\right]
\ee
where $\xi=(D-2)/(2(D-1))=3/8$. As shown in \cite{Hofmann:2003gk}, the conformally coupled scalar will have the profile $\phi=X^{-3/4}(y)B(x^{\mu})$, for $X(y)=Y(y)$ in the line element as written in  \eqref{eq:5dans}. 
Let us write the 5D metric proposal suggested in \cite{Giudice:2016yja} \footnote{Due to the conformal coupling to gravity, adding a bulk CC and using a linear dilaton as in \cite{Giudice:2016yja} will not generate the CW metric, or also later \eqref{eq:xcos}. Since we are interested with the 4D physics, we remain agnostic about the UV completion that generates the action \eqref{eq:Sconformal}}.:
\be
ds^2=e^{\frac{4k|y|}{3}}\left(\eta_{\mu \nu}dx^{\mu}dx^{\nu}+dy^2\right)=\left(\frac{z}{z_0}\right)^2\eta_{\mu \nu}dx^{\mu}dx^{\nu}+dz^2 \label{CWmetric}
\ee
where $z_0\leq z\leq z_{\pi}$. Substituting the CW metric \eqref{CWmetric} the following Ricci curvature and volume element are:
\bea
\mathcal{R}_5=-\frac{16}{3}\left(\frac{k z_0}{z}\right)^2\\
\sqrt{-g_5}=\left(\frac{z}{z_0}\right)^4
\eea
Notice that the numerical coefficient in $\mathcal{R}_5=-16/3=-2/\xi$.  
The action after canonically normalizing the 4D field is:
\be
\mathcal{S}=-\int_0^{\pi R}dy\int d^4x\left\{(\partial_{\mu}\phi)^2+\frac{X^2}{Y^{1/2}}\left(\partial_y \frac{\phi}{X^{1/2}Y^{1/4}}\right)^2+\xi X\mathcal{R}_5\phi^2\right\}
\ee
In any choice of coordinates, substituting \eqref{CWmetric} gives a constant negative mass term due to the conformal coupling:
\be
\mathcal{S}=-\int_0^{\pi R}dy\int d^4x\left\{(\partial_{\mu}\phi)^2+\frac{X^2}{Y^{1/2}}\left(\partial_y \frac{\phi}{X^{1/2}Y^{1/4}}\right)^2-2k^2\phi^2\right\}
\ee
In general, such a term should not necessarily worry us, as this is a negatively curved space-time. However, upon discretization, the 4D metric is Minkowski. 
In such case we shall still get the same expressions for the charges $q_j$ as in \eqref{eq:5dparams}. But, the masses $m^2_{a_j}$ in \eqref{eq:5dparams}, are shifted by a negative constant $-2k^2$. 
\be
{\mathcal S} = \int d^4 x \left[ \sum_{j=0}^N  (\partial_\mu \phi_j)^2 + \sum_{j=0}^{N-1} m_j^2 \left( \phi_j- q_j \phi_{j+1} \right)^2 -2k^2\phi_j^2\right] 
\ee
\be
m_j^2 \equiv \frac{N^2\, X_j}{\pi^2 R^2\, Y_j} \, , ~~~q_j \equiv \frac{X_j^{1/2}Y_j^{1/4}}{X_{j+1}^{1/2}Y_{j+1}^{1/4}} ~.
\label{results}
\ee

Hence, in the 4D picture, all $m_{a_j}^2$ will be shifted by a negative constant. In particular, the massless mode becomes tachyonic, with mass $m_{a_0}^2=-2k^2$.\footnote{In principle, there can be additional negative mass modes depending on $k$ and $R$, the size of the extra dimension.} The tachyon arises due to the negative curvature of the 5D manifold. In RS, we get a similar negative term, but it will be site dependent, since:
\be
\xi X\mathcal{R}_5\phi^2=-\frac{15}{2}\hat k^2e^{2 \hat k z}\phi^2
\ee

Of course one can consider non-minimal coupling to the 5D Ricci scalar that is not conformal and with the opposite sign, such that it does give a positive mass term.\footnote{It may be that such a change of sign simply means moving the tachyon to the gravitational part of the lagrangian. Let's ignore that for the moment.}  But it is not clear what we gain from such a coupling that is not already encapsulated in simply adding a mass term to the 5D scalar, since the conformal symmetry is lost.

Hence, let us consider adding an explicit 5D mass term $M^2\phi^2/2$. This of course breaks the conformal symmetry. In such case, after canonically normalizing the 4D kinetic term we will have:
\be
\mathcal{S}=-\int_0^{\pi R}dy\int d^4x\left\{(\partial_{\mu}\phi)^2+\frac{X^2}{Y^{1/2}}\left(\partial_y \frac{\phi}{X^{1/2}Y^{1/4}}\right)^2+X(M^2+\xi\mathcal{R}_5)\phi^2\right\}
\ee
Since in RS the Ricci scalar is constant, the tachyon can be exactly cancelled, or become massive with $M^2\geq-\xi \mathcal{R}_5=\frac{15}{2}\hat k^2$ upon discretization. Hence, we achieve an explicit breaking at the price of adding another mass scale $M$ to the game. Of course, once we add an explicit 5D mass term, there is no motivation to consider the conformal coupling as the conformal symmetry is lost.

Back to the CW case with minimal coupling to gravity let us consider the 5D mass term.
The contribution of the 5D mass term will be site dependent, according to the value of $X_j$. Generically, there will be no massless mode.
To maintain the massless mode, with minimal coupling to gravity, one has to choose specifically that $M^2=\alpha (q)m^2_j=\alpha(q)(N/\pi R)^2$, where $\alpha(q)$ is some function of $q$ that guarantees the vanishing of the determinant.
For example, in the case of $N=2$, we get $\alpha=q^{-4/3}(1+q^{2/3})$.
With the conformal coupling to gravity, the parameter $\alpha$ becomes dependent on $k$ as well in some complicated expression, $\alpha=\alpha(q,k)$.

To summarize, two possible ways to uplift the massless mode from 5D are conformal coupling to gravity or adding a 5D mass term.
The 5D mass term generates site-dependent mass terms in 4D, but with a judicious choice of parameters, a massless mode can be maintained.
The conformal coupling to gravity makes the CW massless mode tachyonic after discretization. Adding a 5D mass term can remove the tachyonic instability, restoring zero mass for specific values of the parameters $q,k$, or generate a positive 4D mass term.

\subsection{Positively Curved 5D Manifold}
The tachyonic generation of mass is due to the negatively curved nature of the 5D manifold. Considering a positively curved 5D spacetime with constant positive curvature, will give a positive mass term to the massless mode while still maintaining conformal symmetry in 5D.
Considering again the general ansatz \cite{Giudice:2016yja}:
\be
ds^2=X(y)\eta_{\mu \nu}dx^{\mu}dx^{\nu}+Y(y)dy^2\equiv e^{2f(y)}\left(\eta_{\mu \nu}dx^{\mu}dx^{\nu}+dy^2\right)=e^{2g(z)}\eta_{\mu \nu}dx^{\mu}dx^{\nu}+dz^2. \label{CWmetric2}
\ee
 with the choice $X(y)=Y(y)\equiv e^{2f(y)}$. The action after canonically normalizing the 4D field is:
\be
\mathcal{S}=-\int_0^{\pi R}dy\int d^4x\left\{(\partial_{\mu}\phi)^2+\frac{X^2}{Y^{1/2}}\left(\partial_y \frac{\phi}{X^{1/2}Y^{1/4}}\right)^2+\xi X\mathcal{R}_5\phi^2\right\}
\ee
Conformal symmetry is guaranteed if $\xi=3/8$.
We would like to get $X \mathcal{R}_5\equiv e^{2f(y)} \mathcal{R}_5\equiv e^{2g(z)} \mathcal{R}_5=k^2$, for some positive real $k$, ensuring that the gravity induced mass term upon discretization is positive.
Working with the $y$ coordinates, we demand:
\bea
X \mathcal{R}_5=-4\left[2f''(y)+3f'(y)^2\right]=4k^2\\
\Rightarrow X=\cos^{4/3}\left(\frac{\sqrt{3}}{2}k y\right)
\label{eq:xcos}
\eea
The metric is well-defined, provided that $k< 1/\sqrt{3}R$, where $\pi R$ is the size of the extra dimension \footnote{Using the $z$ coordinate $X$ is some hypergeometric function.}. 
\be
\mathcal{S}=-\int_0^{\pi R}dy\int d^4x\left\{(\partial_{\mu}\phi)^2+\frac{X^2}{Y^{1/2}}\left(\partial_y \frac{\phi}{X^{1/2}Y^{1/4}}\right)^2+4\xi k^2\phi^2\right\}
\ee
The discretization proceeds in the same manner as before. For minimal coupling to gravity $\xi\rightarrow 0$, we have a massless mode, the mass parameter is unchanged but the charges now become site-dependent:
\be
m_j^2=\frac{N^2 X_j}{\pi^2R^2Y_j},\quad q_j=\frac{X_j^{1/2}Y_j^{1/4}}{X_{j+1}^{1/2}Y_{j+1}^{1/4}}=\frac{\cos \left(\frac{\sqrt{3}}{2}k j a\right)}{\cos \left(\frac{\sqrt{3}}{2}k (j+1) a\right)},\quad a=\frac{\pi R}{N}
\label{eq:5dconformal}
\ee
With conformal coupling to gravity, $\xi=3/8$, all CW massive modes are shifted by a positive constant $4\xi k^2=\frac{3}{2}k^2$, and similarly, the massless CW goldstone boson gets a mass $m_{a_0}^2=\frac{3}{2}k^2$ and the final $U(1)$ symmetry is broken. Given that the conformal coupling only introduces a fixed mass term for all sites, $V\supset 4\xi k^2 \phi_j^2= 3/2 k^2\phi_j^2$, it only shifts the masses as written above, and it does not change the $\mathcal{O}_{ji}$, where ${i,j}\in [0,N]$. This is because given a system of eigenvalues and eigenvectors, adding a matrix proportional to the identity, only shifts the eigenvalues by the constant of proportionality, and keeps the same eigenvectors. The outcome is a CW model where the breaking of the $U(1)_{CW}$ is due to the conformal coupling to gravity, masses $\tilde m^2_{a_k}=\frac{3}{2}k^2+m^2_{a_k}$ where $m^2_{a_k}$ are the CW massive modes in the absence of conformal coupling, and with site-dependent charges $q_j$ as described in the appendix. The massless mode is:
\be
\mathcal{O}_{j0}=\mathcal{N}_0\left\{1,\frac{1}{q_1},\frac{1}{q_1q_2},\frac{1}{q_1q_2q_3}\cdots,\frac{1}{\Pi_i q_i}\right\}
\ee
Substituting the $q_j$s, specifically for the massless mode, one gets a simple expression of
\be
\mathcal{O}_{j0}=\tilde{\mathcal{N}}\cos\left(\frac{\sqrt{3}}{2}k a\right) \left\{\cos\left(\frac{\sqrt{3}}{2}k a\right),\cos \left(\frac{\sqrt{3}}{2}\,2k a\right),\cos\left(\frac{\sqrt{3}}{2}3k a\right),\cdots,\cos\left(\frac{\sqrt{3}}{2}k(N+1) a\right)\right\}
\ee
So while the mass gap is diminished as all modes have masses $m^2\sim k^2$, the special functional dependence of the original massless mode $\mathcal{O}_{j0}\sim \frac{1}{\Pi_i q_i}$ remains. However, due to the 'telescopic' nature of each $q_j$ we are left with a massless mode that depends rather democratically on all sites.


Before we conclude, in the above analysis, we have not taken into account effects caused by putting branes in the extra dimension when the continuum limit is discussed, as in \cite{Giudice:2016yja, Hambye:2016qkf, Craig:2017cda, Cox:2012ee}. Such construction can have interesting consequences, and we defer such investigation to future work.  
 
 \section{Conclusions}
 In this note, we have attempted to generalize the clockwork idea in several directions, while keeping the original notion of natural generation of hierarchy in a theory whose fundamental parameters are of similar size. Or, in mundane terms, getting an exponential hierarchy, at the price of considering $N+1$ fields. 
 From a lattice point of view, we have demonstrated that coupling beyond nearest neighbors leads to enhanced symmetry, depending on the number of neighbors each site couples to. This is interpreted as higher derivative terms from the extra-dimensional point of view.
 If we allow all possible number of neighbors interactions, the massive eigenstates are shifted considerably, but a massless mode and a residual $U(1)$ remain in tact. The masses of the axions in these constructions are modified and could come close to the mass of the radial modes. We have further generalized the clockwork to global $O(N)$ models.
 
 Generalization to SUGRA is a delicate issue. If we use spurions or stabilizers, the SUSY breaking either has to be as large as the clockwork term for viable phenomenology, or more ingredients have to be added, such as further uplifting of $N$ flat directions, or having the same energy scale for the clockwork and the CC. Simple successful CW SUGRA proceeds either by tuning $w_0$ in the superpotential to cancel the spurions/stabilizers contributions, or by discarding such fields and building a manifestly shift symmetric superpotential. We have also demonstrated how to break SUSY spontaneously and simultaneously conserve or break the CW symmetry for model building purposes. A particularly interesting result, is the fact that the CW symmetry actually dictates the SUSY breaking direction, if one does not want to change the field content of the theory. This direction is exactly the CW massless mode, that gets a finite mass, and neither SUSY nor CW are left in tact. Finally, in the mass basis, a generic SUSY breaking direction which is not the massless mode is available by trivial solution of the extremum equations. 

Conformal coupling of the CW metric in 5D, makes the massless mode tachyonic upon discretization. This is a generic property of negatively curved 5D manifolds once we canonically normalize the kinetic terms of the scalar field and discretize.
A 5D mass term, generically uplifts the massless mode, though for certain value of parameters the masslessness can be restored. This conclusion is valid also in the presence of conformal coupling to gravity in negatively curved 5D manifolds. Finally, with a positively curved $5D$ manifold, one can generate mass for the CW massless mode, but then the charges of the CW scalars become site-dependent.
All in all, it seems we have only started to unravel the various possibilities of the clockwork mechanism.

\section*{Acknowledgements}
I thank Clemens Wieck for many useful comments on the manuscript. I also thank Martin Einhorn and Yevgeny Katz for useful discussions.

\section*{Appendix: Clockwork with various charges}
Let us consider only nearest neighbors interaction, but with a different charge, $q_i$ at each site.
In such a case, if $q_i>3$ we get a nonrenormalizeable theory, and we have to divide by some mass scale $\Lambda^{q_i-3}$.
Thus the general potential looks like:
 \be
\label{eq:general}
V(\phi_j)=\sum_{j=0}^N\left(-\tilde m^2\phi_j^{\dagger}\phi_j+\frac{\lambda}{4}|\phi_j^{\dagger}\phi_j|^2\right)+\sum_{j=0}^{N-1}\left(\epsilon \frac{\phi_j^{\dagger}\phi_{j+1}^{q_j}}{\Lambda^{q_j-3}}+h.c\right)
\ee
We implicitly assume that $\Lambda$ is \textit{larger than any other energy scale of the problem}.
Let us check when can we continue with the separation between the radial and axial modes.
We would like to give each radial field an approximate vev as before of $\langle |\phi_j|^2\rangle= f^2\equiv 2\tilde m^2/\lambda \, ,\forall j$, as well as $f<\Lambda$. For this to happen, we need:
\be
\epsilon\frac{f^{1+q_j}}{\Lambda^{q_j-3}}\ll \frac\lambda f^4 \Rightarrow \epsilon \ll \lambda \left(\frac{\Lambda}{f}\right)^{q_j-3}
\ee
Thus, for $q>3$, the desired hierarchy is easier to fulfill than the original CW. This will also be true when we couple beyond nearest neighbors. The masses of the radial modes are negligibly shifted to $m_{r_k}^2\sim f^2(1+\epsilon(f/\Lambda)^{q_k-3})\sim f^2$.

Below the breaking scale $\sqrt{\lambda}f$, we have a theory of $N+1$ goldstone bosons with $U_j=e^{i \pi_j(x)/f}$ and $j=0,\cdots N$: 
\be
\mathcal{L}=-\sum_{j=0}^{N}f^2\partial U_j^{\dagger}\partial U_j+\frac{m^2}{2}\frac{f^{q_j-1}}{\Lambda^{q_j-3}}\sum_{j=0}^{N-1}\left( U^{\dagger}_jU^{q_j}_{j+1}+h.c\right)
\ee
with $m^2=2\epsilon f^2$.
The CW massless mode will now have the following eigenvector:
\be
\mathcal{O}_{j0}=\mathcal{N}_0\left\{1,\frac{1}{q_1},\frac{1}{q_1q_2},\frac{1}{q_1q_2q_3}\cdots,\frac{1}{\Pi_i q_i}\right\}
\ee
The masses of the axial modes behave as $m_{a_k}^2\sim \epsilon f^2(f/\Lambda)^{q_k-3}$. The exact diagonalization is straightforward.

\end{document}